\documentclass[%
 reprint,
 superscriptaddress,
 amsmath,amssymb,
 aps,
prab,
]{revtex4-2}

\usepackage[colorlinks,linkcolor=red,anchorcolor=green,citecolor=blue,urlcolor=magenta,filecolor=cyan]{hyperref}
\usepackage{caption,siunitx} 
\usepackage{mathrsfs}
\usepackage{graphicx}
\usepackage{dcolumn}
\usepackage{bm}

\begin{document}

\preprint{APS/123-QED}

\title{Symplectic tracking through curved three dimensional fields by a method of generating functions}

\author{Jie Li}
\affiliation{State Key Laboratory of Nuclear Physics and Technology, and Key Laboratory of HEDP of the Ministry of Education, CAPT, Peking University, Beijing, 100871, China}

\author{Kedong Wang}
\email{wangkd@pku.edu.cn }
\affiliation{State Key Laboratory of Nuclear Physics and Technology, and Key Laboratory of HEDP of the Ministry of Education, CAPT, Peking University, Beijing, 100871, China}
\affiliation{Institute of Guangdong Laser Plasma Technology, Baiyun,Guangzhou,510540, China}

\author{Kai Wang}
\author{Xu Zhang}
\affiliation{State Key Laboratory of Nuclear Physics and Technology, and Key Laboratory of HEDP of the Ministry of Education, CAPT, Peking University, Beijing, 100871, China}

\author{Xueqing Yan}
\author{Kun Zhu}
\email{zhukun@pku.edu.cn }
\affiliation{State Key Laboratory of Nuclear Physics and Technology, and Key Laboratory of HEDP of the Ministry of Education, CAPT, Peking University, Beijing, 100871, China}
\affiliation{Institute of Guangdong Laser Plasma Technology, Baiyun,Guangzhou,510540, China}

\date{\today}

\date{\today}

\begin{abstract}
Symplectic integrator plays a pivotal role in the long-term tracking of charged particles within accelerators.
To get symplectic maps in accurate simulation of single-particle trajectories, two key components are addressed: precise analytical expressions for arbitrary electromagnetic fields and a robust treatment of the equations of motion.
In a source-free region, the electromagnetic fields can be decomposed into harmonic functions, applicable to both scalar and vector potentials, encompassing both straight and curved reference trajectories.
These harmonics are constructed according to the boundary surface's field data due to uniqueness theorem.
Finding generating functions to meet the Hamilton-Jacobi equation via a perturbative ansatz, we derive symplectic maps that are independent of the expansion order.
This method yields a direct mapping from initial to final coordinates in a large step, bypassing the transition of intermediate coordinates.
Our developed particle-tracking algorithm translates the field harmonics into beam trajectories, offering a high-resolution integration method in accelerator simulations.
\end{abstract}

\maketitle


\section{\label{sec:Intro}Introduction}
Accurate prediction of particle trajectories in an accelerator beamline is essential for both the design and operation of the accelerator.
This is especially important for a circular machine where particles will revolves for millions of turns, as the effect by any nonlinear fields or complex fringe fields will be magnified.
And the emittance acceptance should be carefully estimated.
Accurate tracking of the exact particle path along a reference trajectory requires a precise and well-defined prior description of the electromagnetic fields.
Standard expressions in simple cases, like hard-edge model, are not applicable to complex magnets.
For example, novel superconducting coils, or complex magnets used for FFAG accelerator, put challenge in calculating particle trajectories~\cite{Wan_prab, FFAG_prab}.
So we need an efficient evaluation of arbitrary static electromagnetic fields before tracking.
An analytical representation of the fields, which can be differentiated and integrated, is more suitable for programming to get the precise result of particle tracking.

For most accelerator elements, the equations of motion can not be solved analytically and must therefore be addressed using numerical integration methods.
In Hamiltonian mechanics, these equations of motion can be equivalently expressed using Hamilton's canonical equations, Poisson brackets, or the Hamilton-Jacobi equation.
Symplecticity, a fundamental property of any Hamiltonian conservative system, ensures the preservation of phase-space structure and is crucial for accurate long-term particle tracking~\cite{Wolski_2012}.
As the system described by the Hamiltonian \(H\) cannot be solved analytically, numerical integration techniques preserving the symplecticity should be employed.
Explicit Runge-Kutta algorithms can be applied to solve Hamilton's canonical equations to obtain particle coordinates and momenta. 
But it require sufficiently small step sizes and do not inherently preserve symplecticity~\cite{Hofer_tracking}.
Poisson brackets naturally extend to Lie algebra method, which is powerful and widely used in particle tracking~\cite{Wu_2003, Symplectic_2018,Skoufaris_2021}.
This includes applications such as the approximation of nonseparable Hamiltonians~\cite{nonseparable_prab} and symplectic multiparticle tracking with self-consistent space-charge simulations~\cite{space_charge_Ji_Qiang}.
For tracking obeying the symplectic condition, each step in these algorithms generate lots of intermediate variables and add complexity to the computation time.
But a Taylor expanded generating function(GF) method can generate a flexible, and symplectic mapping routine for particle tracking in external magnetic fields~\cite{lustfeld1984proper,Hamilton_Jacobi_equation, Symplectic_2011}.
This method is quite general and is based on the solution of the Hamilton-Jacobi equation.
The approach has been successfully used for particle tracking in straight 3D fields for accelerator-driven light sources, allowing a complete parametrization and simplifying tracking studies~\cite{Symplectic_2016}.
It provides a complete set of differential equations to arbitrary orders for the expansion functions.
However, if the magnetic fields in an accelerator are configured to guide particles along a curved trajectory, such as ring accelerators, this application of the generating function method remains unexplored.
To address this, a field representation that satisfies Maxwell's equations must first be established, followed by the development of an efficient method to integrate the equations of motion for particles traversing these fields.

The organization of this paper is as follows: After the introduction, we present the theory of 3D descriptions for magnetic field  using harmonics, and symplectic tracking based on generating functions in \autoref{sec:theory}.
In \autoref{sec:Tracking_result}, we provide the numerical results of our method and illustrate different approaches to addressing the magnetic field.
Finally, conclusions are drawn in \autoref{sec:conclu}.

\section{\label{sec:theory}Theory}
\subsection{\label{subsec:field_repre}Field representation}
Accurate models of the electric and magnetic fields for beamline elements are essential for computing precise particle orbits and high-order transfer maps, which are critical for the performance of accelerators.
In particular, for long straight 2D accelerator magnets, the field quality is efficiently characterized by a set of Fourier coefficients \(a_n, b_n\), known as field harmonics in the polar coordinate system, where \(\rho_0\) is a reference radius.
\begin{equation}
	A_s(\rho, \theta) = \sum_{n=1}^{\infty} \left(\frac{\rho}{\rho_0}\right)^n (a_n \sin n \theta  - b_n \cos n \theta ).
	\label{2D_solu}
\end{equation}

\begin{figure}[htbp]
	\begin{minipage}[t]{0.48\linewidth}
		\centering
		\includegraphics[height=0.25\textwidth]{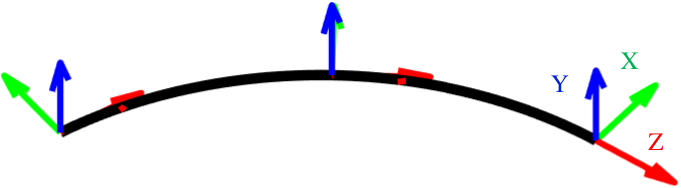}
	\end{minipage}
	\begin{minipage}[t]{0.48\linewidth}
		\centering
		\includegraphics[height=0.25\textwidth]{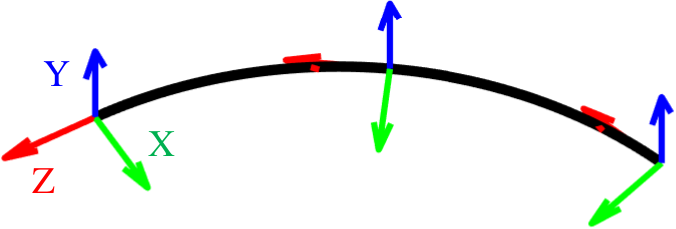}
	\end{minipage}
	\caption{\label{Positive_Negative}The \(x\)-axis (green color) of the local coordinate system is oriented away from the center of curvature for positive \(h\) (left), and toward the center for negative \(h\) (right).}
\end{figure}

In a curved system characterized by curvature \(h\), with invariance along the magnet axis, and using transverse beam coordinates \(x\) and \(y\) centered on the reference trajectory, the longitudinal  magnetic vector potential \(A_s\) is governed by the following equation:
\begin{equation}
	\begin{aligned}
		\frac{\partial ^2A_s}{\partial x^2} & +\frac{\partial ^2A_s}{\partial y^2}+\frac{h}{1+hx}\frac{\partial A_s}{\partial x}-\frac{h^2}{\left( 1+hx \right) ^2}A_s=0,\\
		B_x &=\frac{1}{1+hx}\frac{\partial \left[ \left( 1+hx \right) A_s \right]}{\partial y},\\
		B_y &=\frac{-1}{1+hx}\frac{\partial \left[ \left( 1+hx \right) A_s \right]}{\partial x}.
	\end{aligned}
\label{eq:curved_equ}
\end{equation}
It is noted that the \(x\)-axis of the local coordinate system is oriented away from the center of curvature for positive \(h\) and toward the center for negative \(h\), as illustrated in \autoref{Positive_Negative}.
Specifically, for the dipole field, the solution of \autoref{eq:curved_equ} is expressed as
\begin{equation}
	A_s \sim -x + \frac{hx^2}{2(1 + hx)}.
\end{equation}
Other solutions to \autoref{eq:curved_equ} can be determined through specific analytical techniques~\cite{schnizer2014cylindricalcircularellipticaltoroidal}. By substituting \(A_s\) with \(A_s = \frac{\mathbb{A}}{\sqrt{1 + hx}}\), \autoref{eq:curved_equ} transforms into  
\begin{equation}
	\frac{\partial^2 \mathbb{A}}{\partial x^2} + \frac{\partial^2 \mathbb{A}}{\partial y^2} = \frac{3h^2}{4(1 + hx)^2} \mathbb{A}. 
	\label{eq:omit}
\end{equation}  
This transformed equation isolates the Laplace operator on the left-hand side, suggesting that the preliminary approximate solutions resemble those of \autoref{2D_solu}.
If the curvature is small and the region of interest is very close to the bending axis, the right-hand side of \autoref{eq:omit} can be neglected, and we obtain the following solution:
\begin{equation}
	A_s \approx \frac{\rho^n e^{\text{i}n\theta}}{\sqrt{1 + hx}}.
\end{equation}
To achieve higher accuracy, an iterative method substitutes this initial solution into the right-hand side of \autoref{eq:omit}, refining the left-hand side solution with each iteration to progressively improve the approximation, as follows.
\begin{equation}
		A_s=\frac{\rho ^n e^{\text{i}n\theta}}{\sqrt{1+hx}}\left( 1+\frac{3}{16}\frac{h^2\rho ^2}{n+1} \right).
		\label{detail_approximate}
\end{equation}

Whether in two dimensions or under axisymmetric conditions, these models provide simplified representations of the electromagnetic field distribution, whereas the actual three-dimensional distribution is significantly more complex.
In a region where no net current is enclosed by any closed loop, the magnetic field \(\vec{B}\) is curl-free and can therefore be most simply described in terms of a magnetic scalar potential \(\psi\), \(\vec{B} = \nabla \psi\).
And it must obey the Laplace equation \(\nabla^2 \psi = 0\), which is usually solved numerically via finite element method.

\begin{figure*}[htbp]
	\centering
	\includegraphics[width=13.5cm]{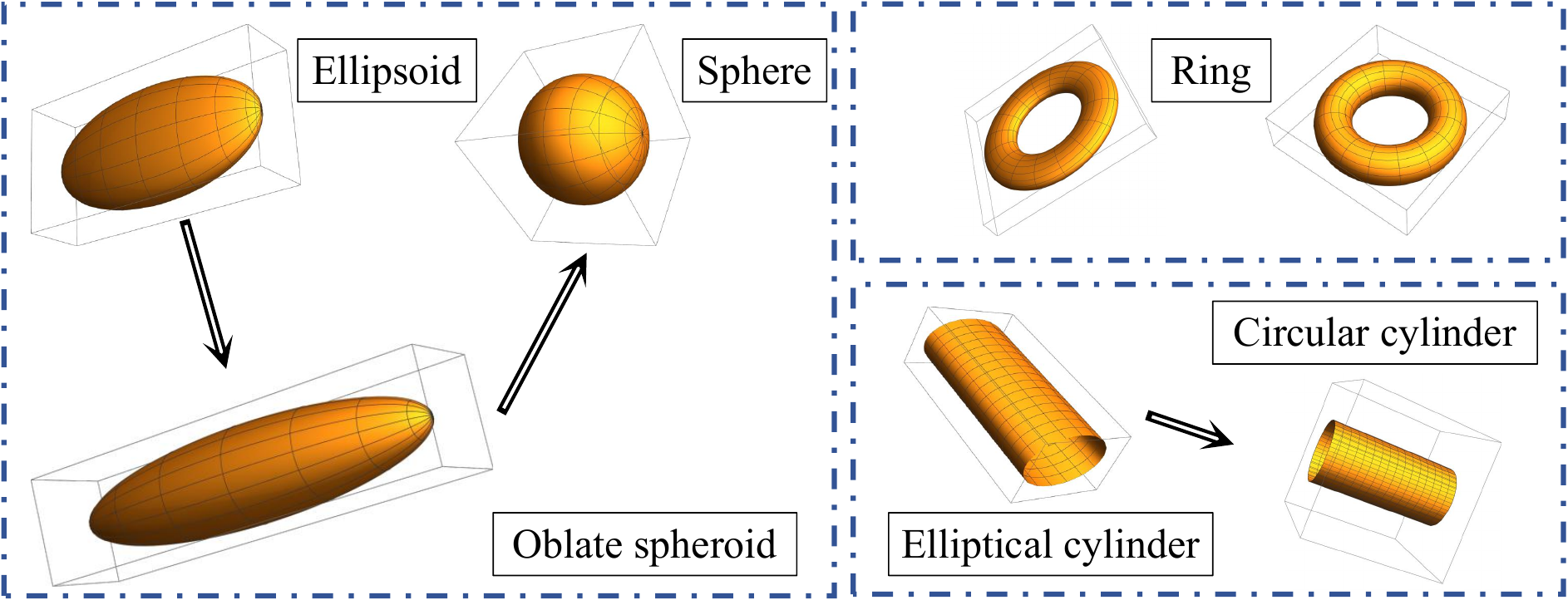}
	\caption{Solutions to Laplace's equation in ring, cylindrical, and spherical domains are typically analyzed within their respective coordinate systems.}
	\label{fig:different_zone}
\end{figure*}

In mathematics, general solutions \(\psi\) satisfying the Laplace equation in certain orthogonal coordinate systems within a specified domain exist as analytic forms, known as harmonic functions.
These magnetic field harmonics form a complete and orthogonal set in the region.
Higher-order terms can be ignored, as it represents rapidly varying magnetic fields that are of minimal practical significance.
Using a set of coefficients to describe the field map, both the field and its derivatives or integrals in the region of interest can be obtained by summing the contributions of the analytical field harmonics.
Mathematicians have extensively studied these solutions in various specific coordinate systems, including ring~\cite{VANMILLIGEN199474}, cylindrical~\cite{Accurate_transfer}, and spherical~\cite{ellipsoidal,spheroidal} domains, as illustrated in \autoref{fig:different_zone}.
Harmonics in cylindrical systems for a straight reference trajectory have been widely discussed in beam physics literature~\cite{Accurate_transfer,Symplectic_2016}.
In this paper, we focus on the harmonics in ring systems for a curved reference trajectory.
Additionally, spherical harmonics are among the most extensively studied harmonic functions, and with appropriate spherical arrangements, they can approximate any shape. 
This paper also explores their applications.

In accelerator physics and beam dynamics, the magnetic vector potential is preferred over the scalar potential, \(\nabla \times \vec{A} = \nabla \psi\).
Here, we choose a gauge where the longitudinal vector potential is set to zero, reducing the variables for trajectory calculations.
The scalar potential is related to the vector potential in a straight coordinate system as
\begin{equation}
	\begin{aligned}
	A_x = \frac{\partial}{\partial y} \left( \int \psi \, dz \right), & \quad  A_y = -\frac{\partial}{\partial x} \left( \int \psi \, dz \right),\\
	  A_z &= 0.
	  \label{scalar_vector_straight}
	\end{aligned}
\end{equation}
In a curved system, the scalar potential is related to the vector potential as
\begin{equation}
	\begin{aligned}
		A_x &= (1 + hx) \frac{\partial}{\partial y} \left( \int \psi \, \text{d}s \right), \\
		A_y &= -(1 + hx) \frac{\partial}{\partial x} \left( \int \psi \, \text{d}s \right), \\
		A_s &= 0.
	\end{aligned}
	\label{curve_A}
\end{equation}

\subsection{\label{subsec:HJfunction}Hamilton-Jacobi equations}

The resolution of mechanical problems using the Hamilton-Jacobi equation adopts a systematic approach rooted in analytical mechanics.
The process begins with the system's Hamiltonian, which, for a particle in an accelerator beamline, is given as~\cite{Beam_Dynamics_in_High_Energy}:
\begin{equation}
	\begin{aligned}
		&H=\frac{\delta}{\beta_0}-(1+hx) \times \left( a_{s} +  \right. \\
		&\left. \sqrt{\left(\delta+\frac{1}{\beta_0}-\frac{q\phi}{cP_0}\right)^2-(p_x-a_x)^2-(p_y-a_y)^2-\frac{1}{\beta_0^2\gamma_0^2}}\right).
	\end{aligned}
\end{equation}
It describes the motion of a particle with energy deviation \(\delta\) in an electrostatic field  \(\phi\) and magnetic fields \(a_x, a_y, a_s\)(both fields are scaled by reference momentum).
The reference particle's velocity, energy, and momentum are represented by \(\beta_0\), \(\gamma_0\), and \(P_0\), respectively.
The longitudinal coordinate \(s\) acts as a time-like parameter, while \(x\) and \(y\) are transverse coordinates with conjugate momenta \(p_x\) and \(p_y\):
\begin{equation}
	\begin{aligned}
		&p_x=\frac{\gamma m_0\beta \frac{\text{d}x}{\text{d}s}+qA_x}{P_0}, \quad p_y=\frac{\gamma m_0\beta \frac{\text{d}y}{\text{d}s}+qA_y}{P_0},\\
		&a_x=\frac{qA_x}{P_0},\quad a_y=\frac{qA_y}{P_0}, \quad a_s=\frac{qA_s}{P_0}.
	\end{aligned}
\end{equation} 
In this paper, we consider a particle with reference energy \(\delta = 0\) traversing an external magnetic field, reducing the problem to two canonical coordinates \(x, y\), and their canonical momenta \(p_x, p_y\).
We assume the dynamical variables remain sufficiently small along the accelerator, enabling the Hamiltonian to be approximated by expanding the square root to a chosen order as follows:
\begin{equation}
	\begin{aligned}
		H & \approx  - \left(1+hx\right)\times\\
		&\left( 1-\frac{1}{2}\left( p_x- a_x \right) ^2-\frac{1}{2}\left( p_y - a_y \right) ^2 + a_s \right),
		\label{equation:approximate}
	\end{aligned}
\end{equation}
which serves as an effective Hamiltonian that maintains symplecticity for particle tracking in this paper.

A specific canonical transformation can be defined such that the transformed Hamiltonian becomes identically zero.
The generating function \(F\) bridges the old coordinates \((x, y)\) and momenta \((p_x, p_y)\), with the new coordinates \((u_x, u_y)\) and momenta \((v_x, v_y)\).
These newly defined coordinates and momenta are constants for each particle, invariant with respect to \(s\) in the motion, though they may differ between particles.
The scalar function \(F(x, y, v_x, v_y)\) satisfies the first-order partial differential equation known as the Hamilton-Jacobi equation:  
\begin{equation}
	\partial _s F(x, y, v_x, v_y) =-H.
	\label{H_J_equation}
\end{equation}
The relationship between the old and new coordinates and momenta is given by: 
\begin{subequations}
	\begin{align}
		p_x = \frac{\partial F}{\partial x},\quad & p_y = \frac{\partial F}{\partial y}; \label{eq:HJ_ab111} \\
		u_x = \frac{\partial F}{\partial v_x},\quad & u_y = \frac{\partial F}{\partial v_y}. \label{eq:HJ_ab222}
	\end{align}
\end{subequations}
\begin{figure}[htbp]
	\begin{minipage}[t]{0.95\linewidth}
		\centering
		\includegraphics[width=0.75\textwidth]{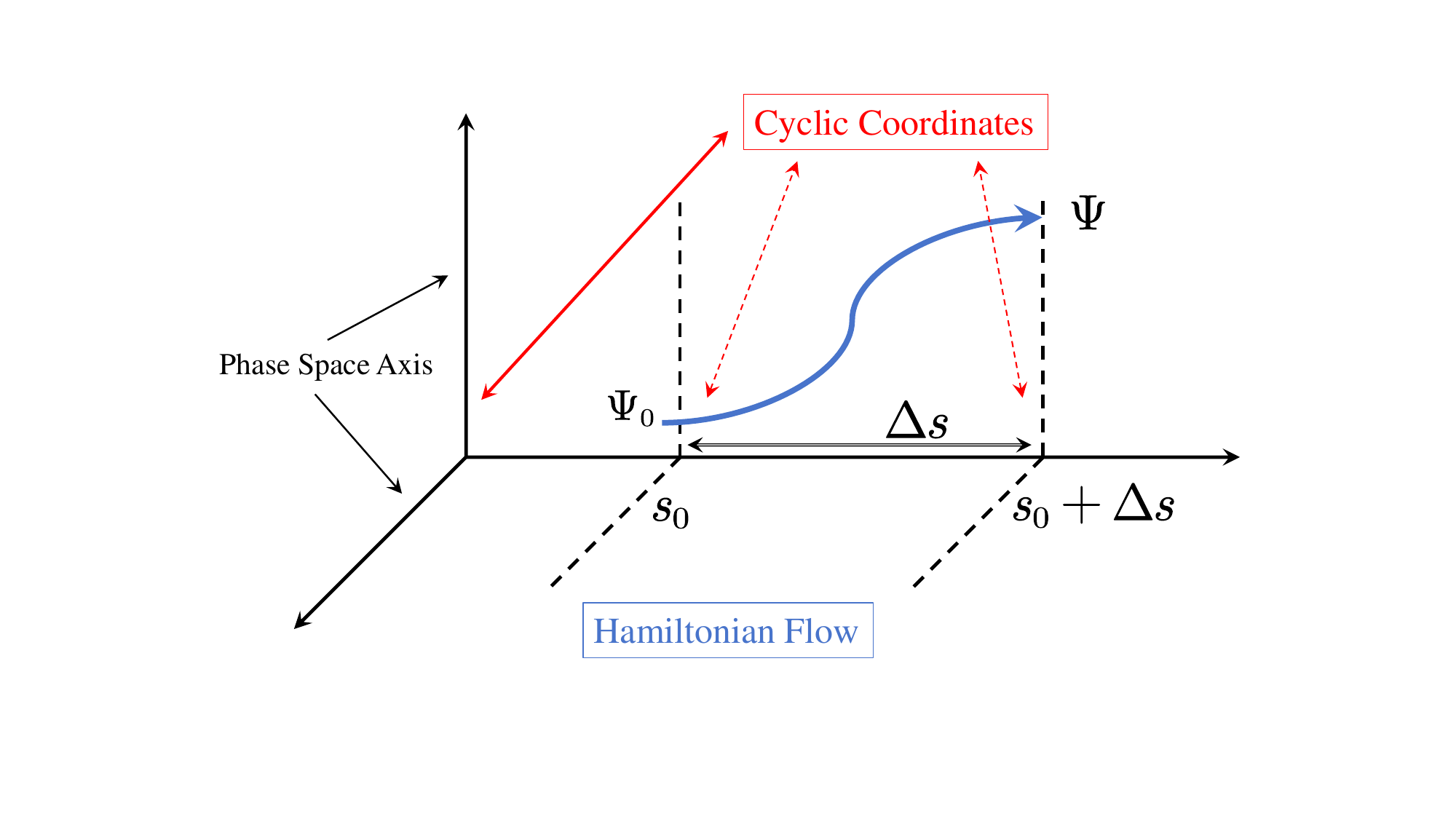}
	\end{minipage}
	\begin{minipage}[t]{0.95\linewidth}
		\centering
		\includegraphics[width=0.75\textwidth]{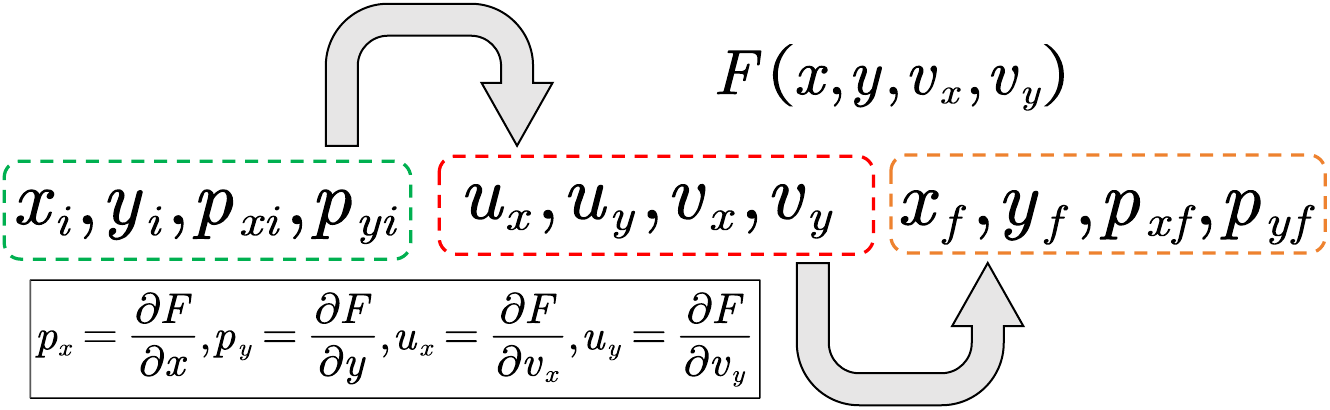}
	\end{minipage}
	\caption{Cyclic coordinates act as invariants in symplectic dynamics, linking initial and final phase-space coordinates via generating functions. }
	\label{fig:GF_particle_Tracking}
\end{figure}
By relating the original coordinates \((x, y)\) and momenta \((p_x, p_y)\) at both ends of a finite step size \(\text{d}s = s_f - s_0\) to the newly defined generalized coordinates \((u_x, u_y)\) and momenta \((v_x, v_y)\), this GF method calculates motions of particles, as illustrated in \autoref{fig:GF_particle_Tracking}.

By substituting equations \autoref{equation:approximate} and \autoref{eq:HJ_ab111} into \autoref{H_J_equation}, we derive the subsequent Hamilton-Jacobi equation:
\begin{equation}
	\begin{aligned}
		&\partial_s F = \left(1+hx\right) \left[ 1-\frac{1}{2}\left( \partial _xF-\varepsilon a_x \right) ^2 \right. \\
		&\left. -\frac{1}{2}\left( \partial _yF-\varepsilon a_y \right) ^2+\varepsilon a_s \right],
		\label{Eq_for_HJ_Eq}
	\end{aligned}
\end{equation}
which involves the first-order partial derivatives of \(F\) with respect to \(s, x, y\).
And we introduce a parameter \( \varepsilon\) for marking magnetic potentials, which should be set to 1 finally.
We adopt a Taylor series expansion of the generating function as an ansatz, of the following form:
\begin{equation}
	F= \sideset{}{'}\sum_{ijkl=0}^{ijkl=\infty}{f_{ijkl}v_{x}^{i}v_{y}^{j}h^k\varepsilon ^l} + v_xx + v_yy,
	\label{ansatz_formula} 
\end{equation}
where \(f_{1, 0 ,0 ,0 } = x\) and \(f_{0, 1,0 ,0 } = y\) are excluded from the summation.
The \(f_{i, j, k, l}\) coefficients are functions of the local variables \(x, y, s\), and the expansion order is given by \(i + j + k +l\).
By substituting \autoref{ansatz_formula} into \autoref{Eq_for_HJ_Eq} and considering some of the lowest-order  \(f_{i, j, k, l}\) terms, we can recursively obtain higher-order terms of  \(f_{i, j, k, l}\). 
This property arises from the presence of the quadratic terms on the right-hand side of \autoref{Eq_for_HJ_Eq}.
The use of parameter \(\varepsilon\) just facilitates the track of the power of the potential functions \(a_x, a_y, a_s\) for coefficient \(f_{i, j, k,l}\),  and therefore can be set to \( \varepsilon = 1\), without loss of generality.
It is noted that the introduction of the parameter \( \varepsilon\) enables the formulation of a recurrence relation for coefficients \(f_{i, j, k,l}\).

Once we have determined the coefficients of generating function \(f_{i, j, k,l}\),  \autoref{eq:HJ_ab111} and \autoref{eq:HJ_ab222} will yield the following relationships:
\begin{equation}
	\begin{aligned}
		p_x &= \sideset{}{'}\sum_{ijkl=0}^{ijkl=\infty}{ \frac{\partial f_{ijkl}}{\partial x}v_{x}^{i}v_{y}^{j}h^k\varepsilon ^l}+v_x,\\
		p_y &= \sideset{}{'}\sum_{ijkl=0}^{ijkl=\infty}{\frac{\partial f_{ijkl}}{\partial y}v_{x}^{i}v_{y}^{j}h^k\varepsilon ^l}+v_y.
		\label{eq:HJ_333}
	\end{aligned}
\end{equation}
\begin{equation}
	\begin{aligned}
		u_x &= \sideset{}{'}\sum_{ijkl=0}^{ijkl=\infty}{ if_{ijkl}v_{x}^{i-1}v_{y}^{j}h^k\varepsilon ^l }+x,\\
		u_y &= \sideset{}{'}\sum_{ijkl=0}^{ijkl=\infty}{	jf_{ijkl}v_{x}^{i}v_{y}^{j-1}h^k\varepsilon ^l}+y.
		\label{eq:HJ_444}
	\end{aligned}
\end{equation}
The primed summation excludes  \(f_{1000}\) and \(f_{0100}\).
If the condition \(f_{i, j, k,l}|_{s=s_f} \equiv  0\) is satisfied, then the new coordinates \(u_x, u_y\) and momenta \(v_x, v_y\), equals to the old coordinates \( x|_{s = s_f}, y|_{s = s_f} \) and momenta \( p_x|_{s = s_f}, p_y|_{s = s_f} \), respectively.
Here in a static field, the final coordinates \(s_f\) can potentially precede the initial coordinates \(s_0\).
With the given order above for \(F\) and a small step size  of \(|s_0 - s_f|\), the Newton routine to obtain \(p_x|_{s = s_f}\) and \(p_y|_{s = s_f}\) can be employed to solve \autoref{eq:HJ_333} as follows.
\begin{equation}
	\begin{aligned}
		v_x &= p_x|_{s=s_0} - \sum_{ijkl=0}^{ijkl=\infty}{\frac{\partial f_{ijkl}}{\partial x} v_x^i v_y^j h^k \varepsilon^l }, \\
		v_y &= p_y|_{s=s_0} - \sum_{ijkl=0}^{ijkl=\infty}{\frac{\partial f_{ijkl}}{\partial y} v_x^i v_y^j h^k \varepsilon^l}.
	\end{aligned}
\end{equation}
In this process, the initial guess values for \( v_x \) and \( v_y \) can be set to \( p_x|_{s=s_0} \) and \( p_y|_{s=s_0} \), respectively. Typically, only a few iterations are required to converge to the solutions.

For example, in the case of no fields, so-called ``drift case'', we have Hamiltonian
\begin{equation}
 \partial_s F = 1 - \frac{1}{2} (\partial_x F)^2  - \frac{1}{2} (\partial_y F)^2 .
\end{equation}
And the following is non-zero coefficient:
\begin{equation}
		f_{0000} = s-s_f, f_{2000} = -\frac{s-s_f}{2},f_{0200} = -\frac{s-s_f}{2}.
\end{equation}
And the results from \autoref{eq:HJ_333} and \autoref{eq:HJ_444} satisfy the prediction for drift segment.
\begin{equation}
	\begin{aligned}
			 p_x|_{s = s_0} &= v_x = p_x|_{s = s_f},\\
			 p_y|_{s = s_0} &= v_y = p_y|_{s = s_f}.
	\end{aligned}
\end{equation}
\begin{equation}
	\begin{aligned}
	 	u_x &= x|_{s = s_f} = v_x (s_f-s_0) = p_x|_{s = s_0}  (s_f-s_0), \\
	 	u_y &= y|_{s = s_f} = v_y (s_f-s_0) = p_y|_{s = s_0}  (s_f-s_0).
	\end{aligned}
\end{equation}

\section{\label{sec:Tracking_result}Tracking results}

\subsection{\label{sec:Tracking_simply_2D}Tracking through ideal elements}
In the first test, we assess the suitability of the generating function method for tracking motion through ideal hard-edge elements.
And we want to emphasize the importance of precise representation of magnetic field.
In nonlinear dynamical systems, the accurate tracking of dynamical variables for long integration periods is paramount for a thorough understanding of their evolution.

We formalize definitions to characterize the algorithmic implementation of recurrence relations governing the coefficients \(f_{ijkl}\).
By substituting the ansatz and performing a coefficient comparison for the Hamilton-Jacobi equation, we derive the following equations:
\begin{equation}
	\begin{aligned}
		&Z_{\left( ijkl \right)}=\\
		&-\frac{1}{2}\sum_{\alpha =0}^i{\sum_{\beta =0}^j{\sum_{\gamma =0}^k{\sum_{\chi =0}^l{\left[ X_{\left( \alpha ,\beta ,\gamma ,\chi \right)}X_{\left( i-\alpha ,j-\beta ,k-\gamma ,l-\chi \right)} \right. }}}}\\
		&\left.+Y_{\left( \alpha ,\beta ,\gamma ,\chi \right)}Y_{\left( i-\alpha ,j-\beta ,k-\gamma ,l-\chi \right)} \right]\\
		&-\frac{x}{2}\sum_{\alpha =0}^i{\sum_{\beta =0}^j{\sum_{\gamma =0}^{k-1}{\sum_{\chi =0}^l{\left[ X_{\left( \alpha ,\beta ,\gamma ,\chi \right)}X_{\left( i-\alpha ,j-\beta ,k-1-\gamma ,l-\chi \right)}\right. }}}}\\
		&\left.+Y_{\left( \alpha ,\beta ,\gamma ,\chi \right)}Y_{\left( i-\alpha ,j-\beta ,k-1-\gamma ,l-\chi \right)} \right]
	\end{aligned}
\end{equation}
where the definitions are given by:
\begin{equation}
	\begin{aligned}
		Z_{ijkl} := & f_{ijkl;s}-\delta _{i}^0\delta _{j}^0\delta _{k}^0\delta _{l}^1a_s  \\
		&-\delta _{i}^0\delta _{j}^0\delta _{k}^1\delta _{l}^1 x a_s, \\
		X_{ijkl} := &f_{ijkl;x}-\delta _{i}^0\delta _{j}^0\delta _{k}^0\delta _{l}^1a_x, \\
		Y_{ijkl} := & f_{ijkl;y}-\delta _{i}^0\delta _{j}^0\delta _{k}^0\delta _{l}^1a_y.
		\label{some_definition}
	\end{aligned}
\end{equation}
The notation \(f_{ijkl;\star}\) denotes the partial derivatives with respect to the associated variables.
Higher-order \(f_{ijkl}\) is computed by successively differentiating lower-order terms with respect to \(x\) and \(y\) and then integrating them with respect to \(s\).

The magnetic field for ideal hard-edge elements is derived from the solution presented in \autoref{2D_solu}.
In this scenario, only the potential function \( a_s \) exists, while \( a_x \) and \( a_y \) are zero.
In this case, the longitudinal variable \( s \) is decoupled from the transverse variables \( x \) and \( y \), which allows the condition \( f_{i, j, k, l} |_{s=s_f} \equiv 0 \) to be easily satisfied.
Several coefficients, written down explicitly, are
\begin{equation}
	\begin{aligned}
		f_{0000} &= s - s_f, f_{2000} = f_{0200} = -\frac{s-s_f}{2}, f_{0001} = \int a_s ds, \\
		f_{1001} &= \int \left( \int \frac{\partial a_s}{\partial x} \, ds' \right) ds, f_{0101} = \int \left( \int \frac{\partial a_s}{\partial y} \, ds' \right) ds, \\
		f_{0002} &= -\frac{1}{2} \int \left[ \left( \int \frac{\partial a_s}{\partial x} \, ds' \right)^2 + \left( \int \frac{\partial a_s}{\partial y} \, ds' \right)^2 \right] ds, \\
		f_{0003} &= -\int \left( \partial_x f_{0001} \partial_x f_{0002} + \partial_y f_{0001} \partial_y f_{0002} \right) ds, \\
		f_{0010} &=x(s-s_f);f_{2010}=f_{0210}=-\frac{x(s-s_f)}{2};\\
		f_{0011} &=xf_{0001};f_{1011}=xf_{1001};f_{0111}=xf_{0101};\\
		f_{0012} &=xf_{0002};f_{0013}=xf_{0003}.
	\end{aligned}
\end{equation}

According to \autoref{detail_approximate}, the magnetic field for curved quadrupoles, accurate to the second order of \(h\), is given by:
\begin{subequations}
	\begin{align}
	A_s = & \left( x^2-y^2 \right) +\frac{h}{2}\left( xy^2-x^3 \right) \nonumber \\
	&+\frac{h^2}{16}\left( 7x^4-6x^2y^2-y^4 \right),\\
	B_x =& -2y+hxy-\frac{h^2}{4}\left( 3x^2y+y^3 \right), \\
	B_y =& -2x+\frac{h}{2}\left(x^2+y^2 \right) -\frac{h^2}{4}\left(x^3+3xy^2 \right).
	\end{align}
\end{subequations}

\begin{figure}[htbp]
	\begin{minipage}[t]{0.98\linewidth}
		\centering
		\includegraphics[width=4.8cm]{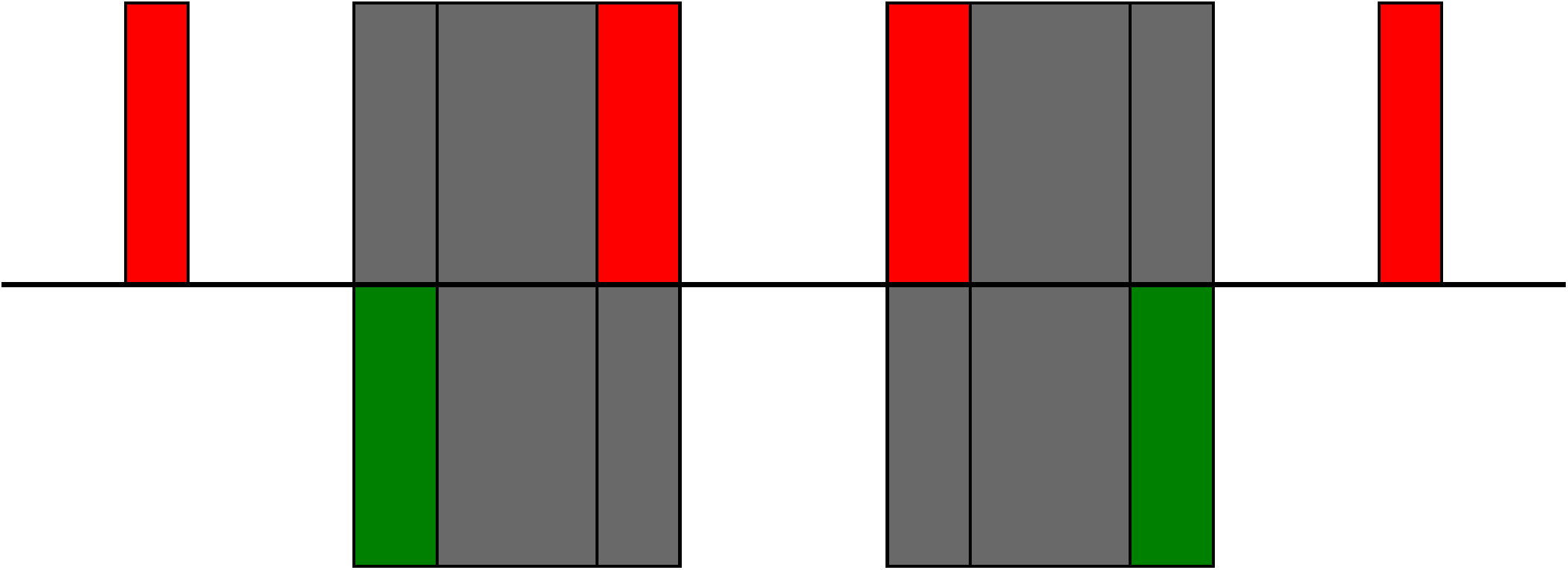}
	\end{minipage}
	\begin{minipage}[t]{0.98\linewidth}
		\centering
		\includegraphics[width=4.8cm]{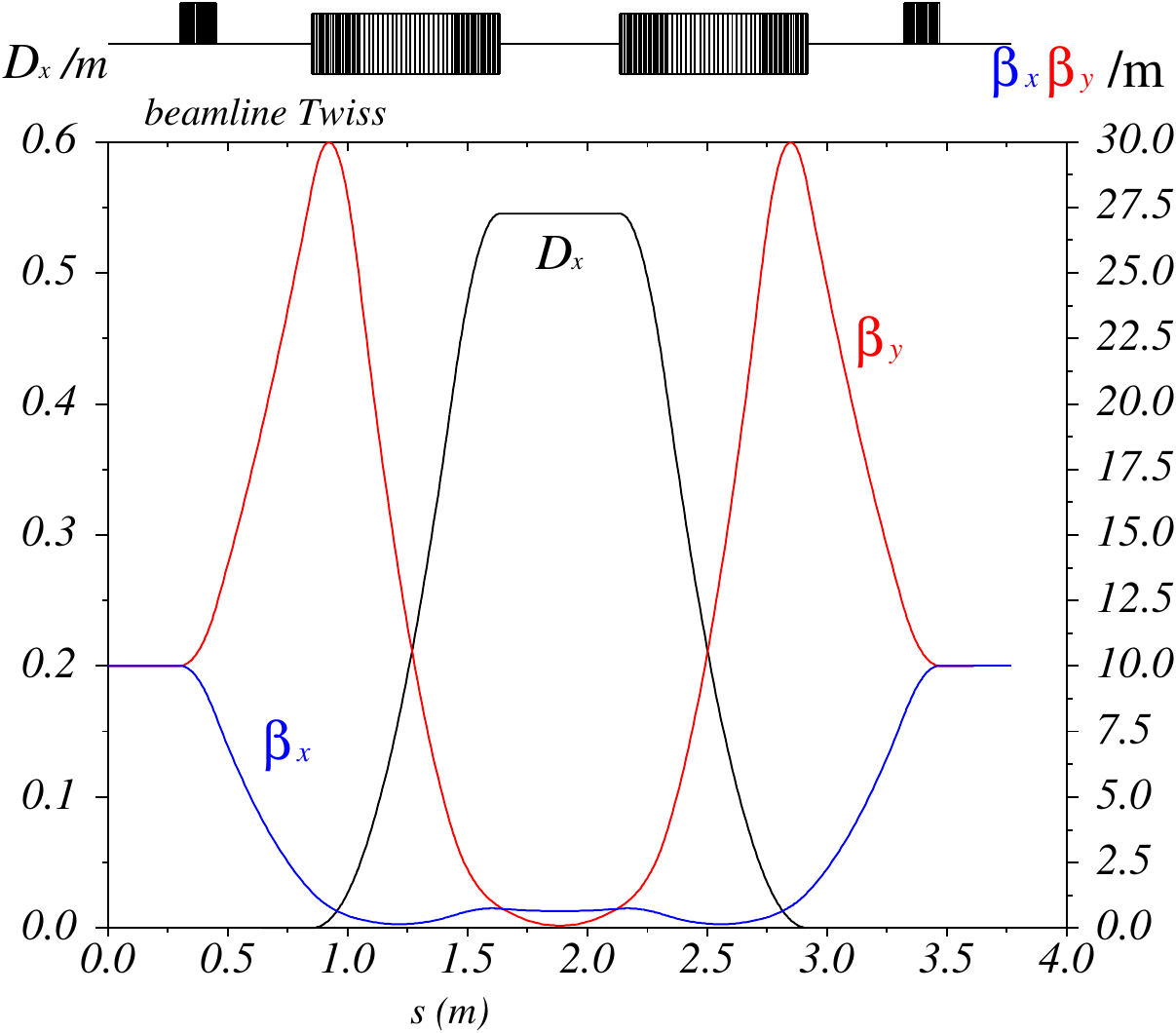}
	\end{minipage}
	\caption{Simple lattice layout with curved quads and dipoles. Top: The half of the symmetric lattice includes a drift, a straight focusing quad, a drift, a curved defocusing quad with a dipole, a dipole, a curved focusing quad with a dipole, and a drift sequentially. Bottom: The evolution of the Twiss parameters for the lattice in MADX.}
	\label{fig:lattice_period}
\end{figure}

A beamline is utilized to demonstrate the significance of accurately representing the magnetic fields of curved elements.
The beamline employs a combined field consisting of curved quadrupoles and dipoles with curvature \(h=\SI{1}{m^{-1}}\), as illustrated in \autoref{fig:lattice_period}, which is a popular design in tumor therapy transfer systems.
We track a single test particle for 3000 turns within the beamline.
The expansion order of \(f_{ijkl}\) is set to \(i \leq 2, j \leq 2, k \leq 2, l \leq 3 \), and the step length is set to \(s_f-s_0=\SI{1}{mm}\).
Further increasing the expansion order does not alter the results much.
We compare 3 different field models for the curved quadrupoles in tracking: linear field, accuracy up to \(h\) and accuracy up to \(h^2\).

\begin{figure}[htbp]
	\begin{minipage}[t]{0.98\linewidth}
		\centering
		\includegraphics[width=5.4cm]{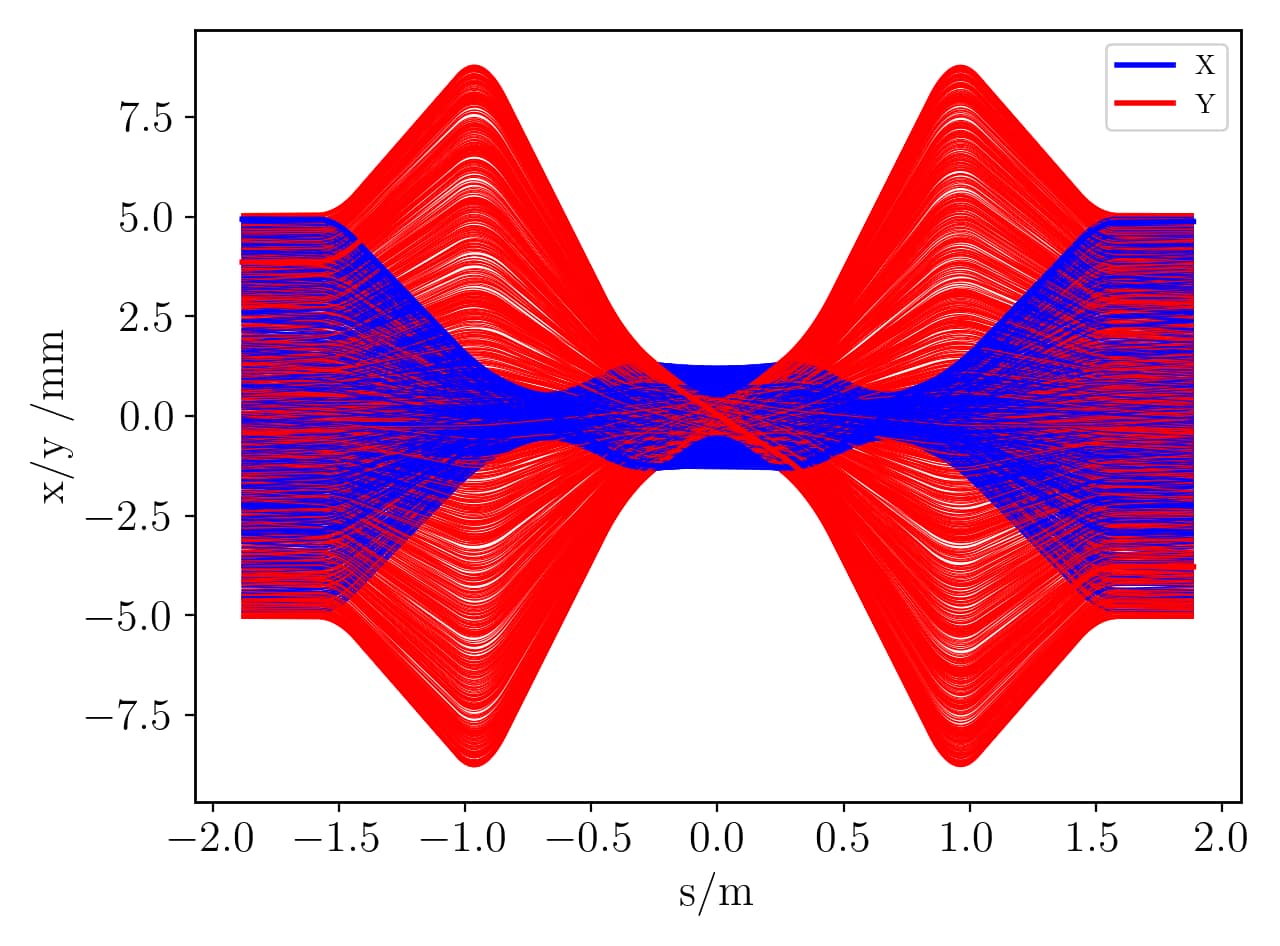}
	\end{minipage}
	\caption{The particle trajectory in the beamline for \(3 \times 10^3\) turns.}
	\label{fig:simple_envelope}
\end{figure}

\begin{figure}[htbp]
	\begin{minipage}[t]{0.98\linewidth}
		\centering
		\includegraphics[width=0.96\textwidth]{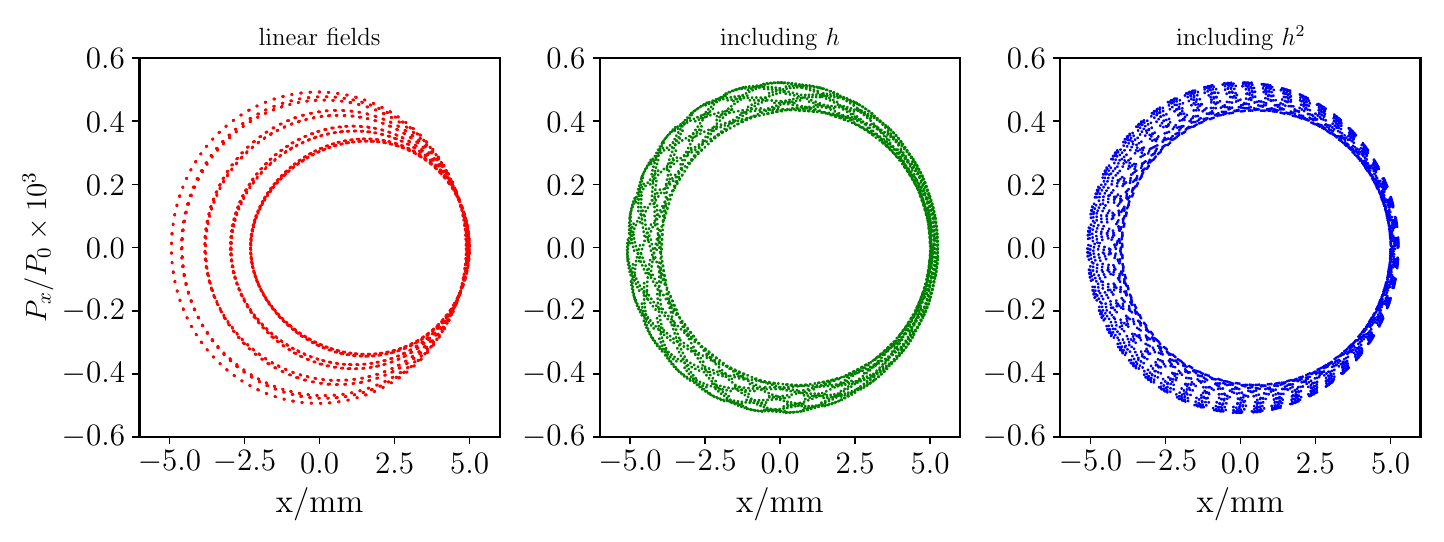}
	\end{minipage}
	\begin{minipage}[t]{0.98\linewidth}
		\centering
		\includegraphics[width=0.96\textwidth]{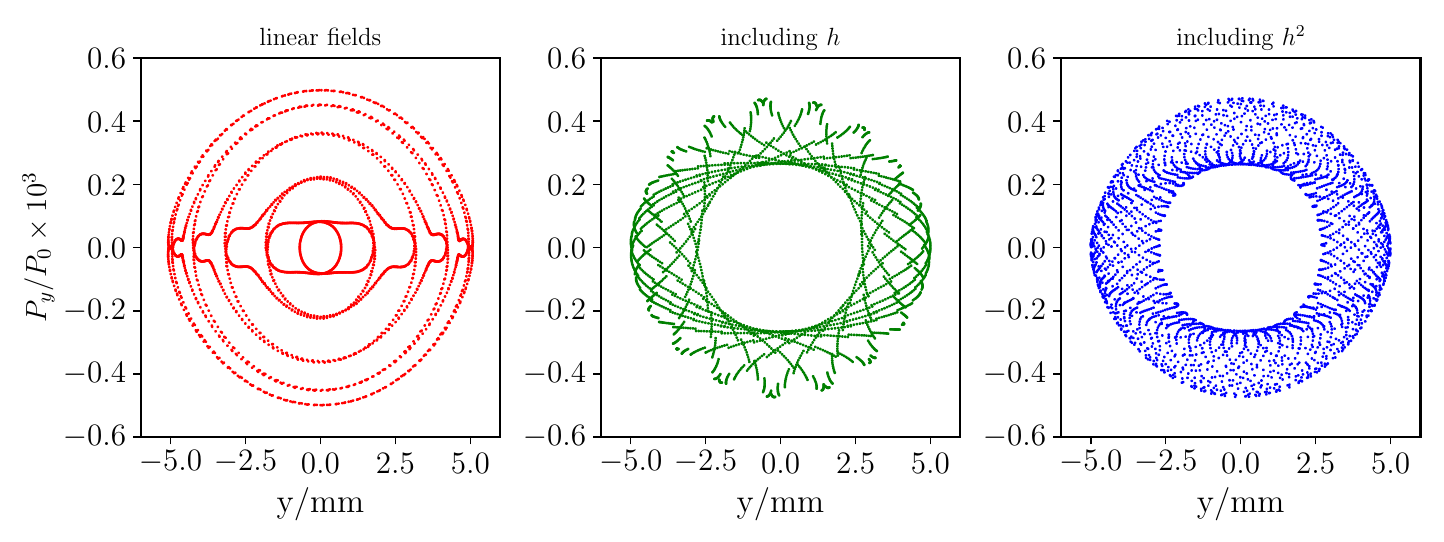}
	\end{minipage}
	\caption{Phase space of one test particle in the beamline for for \(3 \times 10^3\) turns. Top: \(x-p_x\) phase space. Bottom: \(y-p_y\) phase space. The three columns, from left to right,   correspond to different field models with varying magnetic field accuracies: the linear field, accuracy up to \(h\), and accuracy up to \(h^2\).}
	\label{fig:phase_space}
\end{figure}

\begin{figure}[htbp]
	\centering
	\includegraphics[width=7.2cm]{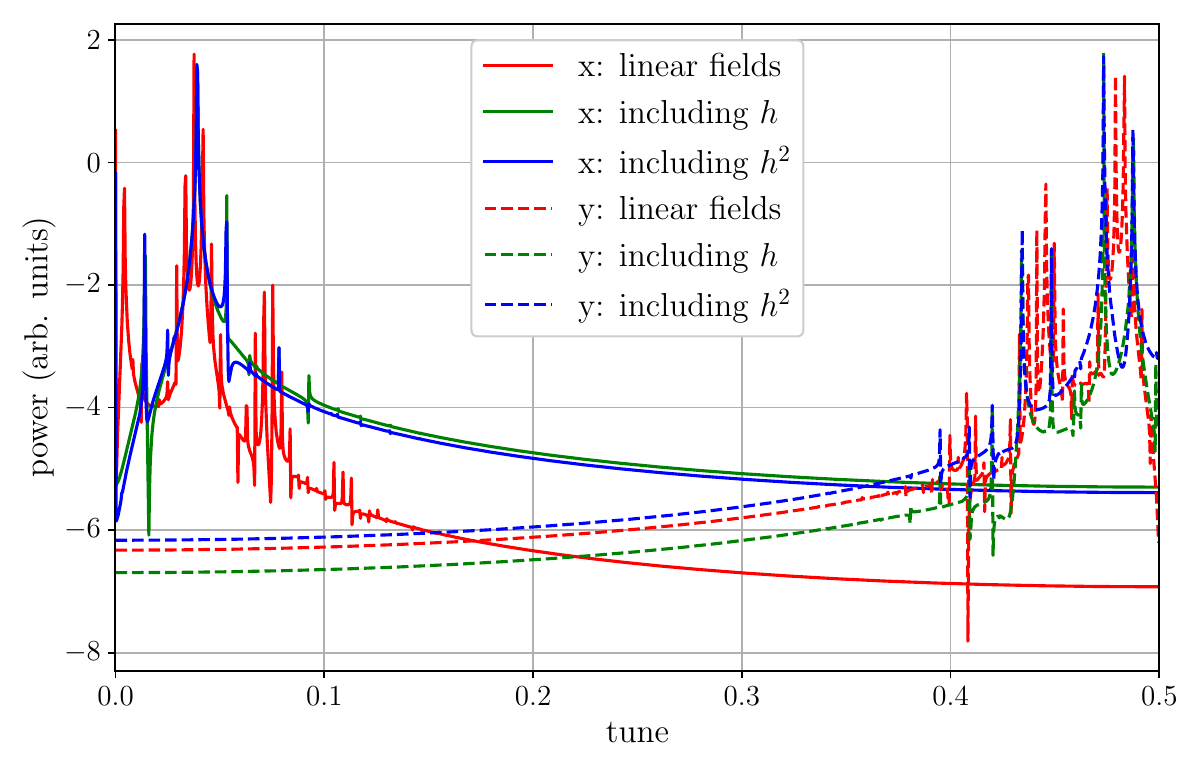}
	\caption{Logarithmic plot of the Fourier transform of the orbit excursion over \(3 \times 10^3\) turns. The peaks correspond to the tunes. Solid lines indicate the results in the x-direction, while dashed lines indicate the results in the y-direction. Different colors represent curved quadrupoles with varying magnetic field accuracies: red for the linear field, green for accuracy up to \(h\), and blue for accuracy up to \(h^2\).}
	\label{fig:tune_fig}
\end{figure}

The particle trajectory of 3000 turns can be seen in \autoref{fig:simple_envelope}, which agrees with the Twiss parameter evolution.
The amplitude of the transverse orbit excursion has reached \SI{8}{mm}, and different field models will have varying effects on long-term tracking. 
And we record its transverse displacements at the entrance of beamline to obtain the phase space, as shown in \autoref{fig:phase_space}.
For different field models of curved quadrupoles, particularly when comparing the linear field model with the model incorporating \(h\) correction, significant differences are observed in the phase space. 
But the phase space characteristics of the field model with \(h\) correction and the one with \(h^2\) correction look similar.
The orbit displacement as a function of turn number is Fourier transformed to determine the tune and the dynamic behavior of the particle, as depicted in \autoref{fig:tune_fig}.
The tune of the linear field model deviates from the other two models, while the latter two exhibit similarities.
This shows the importance of accurate field descriptions.

\subsection{\label{sec:Tracking_bipolar}Tracking in bipolar coordinates}

Prior work \cite{Accurate_transfer} models straight magnet elements (including fringe fields and multipoles) via cylindrical surface data. 
And then \cite{Symplectic_2016} demonstrated symplectic tracking along the straight reference trajectory.
And our paper focuses on magnet curved elements, which needs the field representation of curved beamline elements in ring (or called ``bipolar'') coordinates~\cite{Gambini2018ExpansionOT}.

\begin{figure}[htbp]
	\centering
	\includegraphics[width=3.5cm]{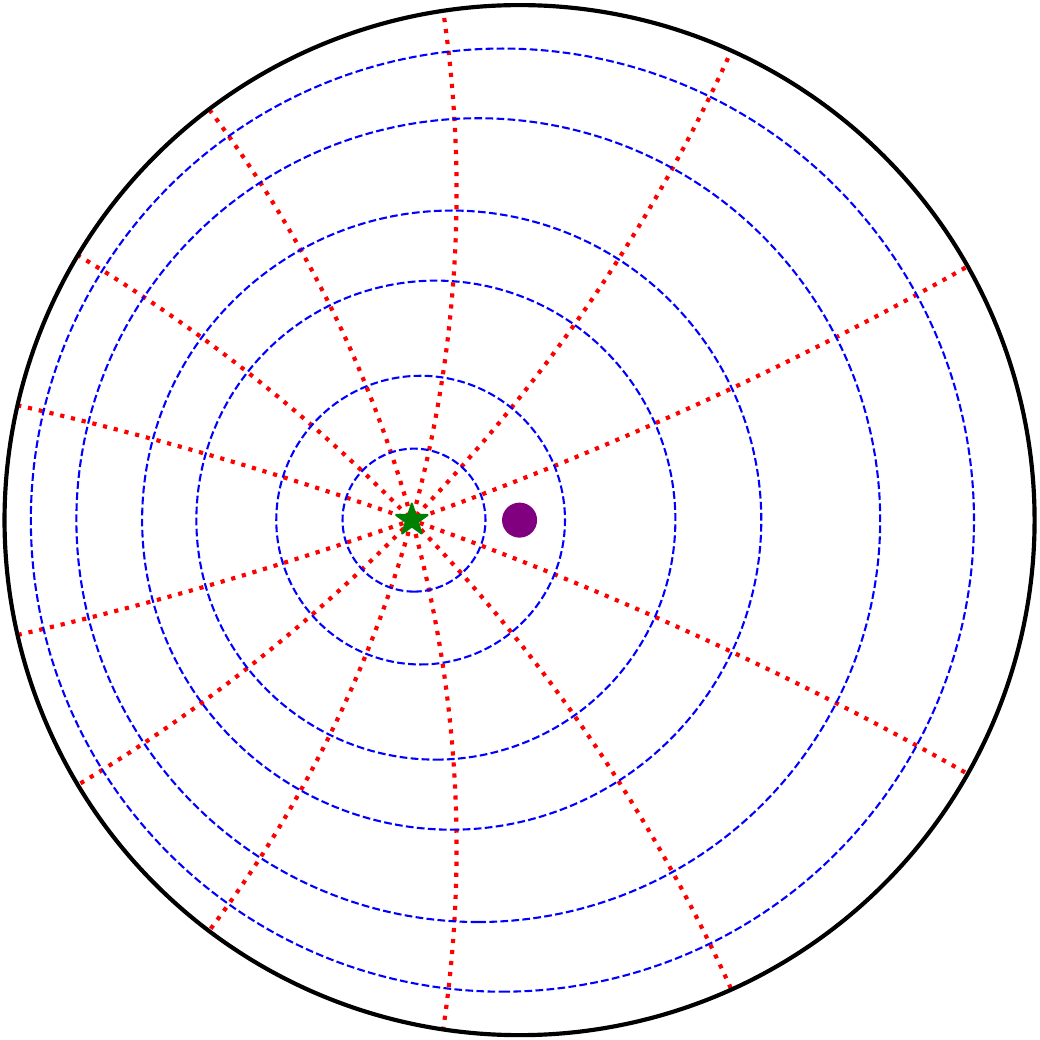}
	\caption{In bipolar coordinates, the red curves represent lines of constant angular variable \(\xi\) ranging from 0 to \(2\pi\), while the blue curves correspond to lines of constant radial variable \(\eta\) within the bore surface (shown in black). Larger values of \(\eta\) correspond to circles of decreasing diameter, and the focus (green pentagram marker, \(\eta =\infty\)) for the bipolar system moves away from the center (purple circular marker, \(x =0, y = 0\) ) of the bore aperture.}
	\label{fig:ring_bipolar}
\end{figure}
The coordinates in the transverse plane for curved reference trajectory inside the magnet's bore are illustrated in \autoref{fig:ring_bipolar}.
The bipolar coordinates \( \eta\) (radial) and \(\xi\) (angular), are related to \(x\) and \(y\) in the beam coordinate system by 
\begin{equation}
	\begin{aligned}
		&\coth\frac{\eta -i\xi}{2} = \frac{1/h +x+iy}{c}, \\
		& x =\frac{c\sinh \eta}{\cosh\eta -\cos \xi}-1/h, \\
		& y =\frac{c\sin \xi}{\cosh \eta -\cos \xi},
	\end{aligned}
\end{equation}
where \( h > 0 \) is the central curvature of the magnet’s bore, and \( c \) is the focus in the bipolar coordinates.
Here, we define a variable \( \text{k}=\cosh(\eta)-\cos(\xi) \).
The unit vectors \(\hat{x}\), \(\hat{y}\), and \(\hat{s} = \hat{x} \times \hat{y} \) in the curvilinear coordinate system for particles correspond to \(\hat{R}\), \(\hat{Z}\), and \(-\hat{\phi}\), respectively, in the cylindrical coordinate system.
We then find formulas concerning partial derivatives:
\begin{equation}
	\begin{aligned}
		\left. \frac{\partial \eta}{\partial y} \right|_x &=\left. \frac{\partial \xi}{\partial x} \right|_y=-\frac{\sin \xi \sinh \eta}{c},\\
		\left. \frac{\partial \eta}{\partial x} \right|_y &= \left. -\frac{\partial \xi}{\partial y} \right|_x=-\frac{\cosh\eta \cos \xi -1}{c},\\
		\left. \frac{\partial x}{\partial \xi} \right|_{\eta} &= \left. \frac{\partial y}{\partial \eta} \right|_{\xi}=-\frac{c\sinh \eta \sin \xi}{\text{k}^2},\\
		\left. \frac{\partial y}{\partial \xi} \right|_{\eta} &= \left. -\frac{\partial x}{\partial \eta} \right|_{\xi}=\frac{c\left( \cosh \eta \cos \xi -1 \right)}{\text{k}^2}.
	\end{aligned}
\end{equation}

In terms of the bipolar coordinates, the scalar potential \(\psi\) can be written as follows~\cite{Symplectic_2018}:
\begin{equation}
	\begin{aligned}
		&\nabla^2\psi=0,\quad\vec{B}=\nabla\psi = \nabla \times \vec{A},\\
		&\psi\left(\eta,\xi,\phi\right)=M_{00}^{\phi}\phi+\sqrt{\cosh(\eta)-\cos(\xi)} \\
		&\sum_{m=-\infty}^\infty\sum_{n=-\infty}^\infty f_{m,n} \frac{Q_{|m|-\frac12}^{|n|}\left(\cosh(\eta)\right)}{Q_{|m|-\frac12}^{|n|}\left(\cosh(\eta_0)\right)} e^{\text{i}n\phi} e^{\text{i}m\xi},
	\end{aligned}
\end{equation}
where the \(f_{m,n}\) are coefficients representing the strength of one harmonic, \(\eta_0\) is the radial bipolar coordinate for the bore, and \(M_{0,0}^\phi\) represents the contribution from the net current enclosed by the curved trajectory, which is not considered in this paper.
And \(Q_{m-\frac{1}{2}}^{n} (\cosh \eta)\) is the associated Legendre function of of the second kind(or called toroidal functions, ring functions), and the derivatives can be written by the recurrence relation.
\begin{equation}
	\begin{aligned}
		\frac{\text{d}Q_{m-\frac{1}{2}}^{n}}{\text{d}\eta}=&\left( m+n-\frac{1}{2} \right) \left( m-n+\frac{1}{2} \right) Q_{m-\frac{1}{2}}^{n-1}-n\frac{Q_{m-\frac{1}{2}}^{n}}{\tanh\eta},\\
		\frac{\text{d}Q_{m-\frac{1}{2}}^{n}}{\text{d}\eta}=&\frac{m-\frac{1}{2}}{\tanh\eta}Q_{m-\frac{1}{2}}^{n}-\frac{m+n-\frac{1}{2}}{\sinh\eta}Q_{m-\frac{3}{2}}^{n}.
	\end{aligned}
\end{equation}
In bipolar coordinates, the vector potential can be
expressed as follows according to \autoref{curve_A} (here, the azimuth direction \(\hat{\phi}\) is opposite to that of \(\hat{s}\)):
\begin{equation}
	\begin{aligned}
		A_{\xi} &= -\sinh \eta \frac{\partial}{\partial \eta} \left( \int \displaystyle{\psi \, \text{d}\phi} \right),\\
		A_{\eta} &= \sinh \eta \frac{\partial}{\partial \xi} \left( \int \displaystyle{\psi \, \text{d}\phi} \right), \\
		A_x &= -\frac{c \sinh \eta}{\text{k}} \frac{\partial}{\partial y} \left( \int \displaystyle{\psi\, \text{d}\phi} \right), \\
		A_y &=  \frac{c \sinh \eta}{\text{k}} \frac{\partial}{\partial x} \left( \int \displaystyle{\psi \, \text{d}\phi} \right),\\
		A_\phi &=0,\quad A_s = 0.
		\label{vector_ring_potential}
	\end{aligned}
\end{equation}

\begin{figure}[htbp]
	\centering
	\includegraphics[width=7cm]{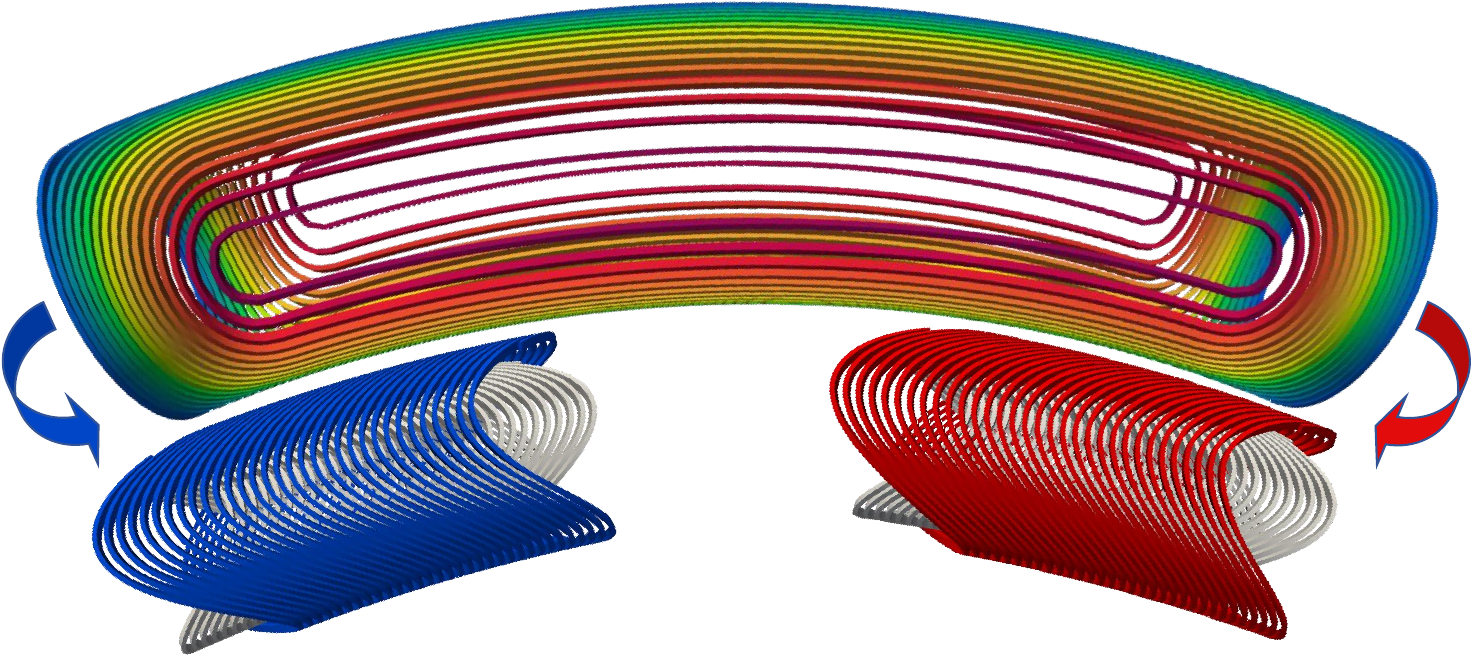}
	\caption{A bending magnet composed of a dipole (DCT coil) with integrated quadrupoles (CCT coils) on each side of the bore provides combined bending and focusing/defocusing for charged particles.}
	\label{fig:coil_displace}
\end{figure}

To illustrate the application of the generating function in bipolar coordinates, we consider a complex magnet configuration, as depicted in \autoref{fig:coil_displace}. 
This configuration comprises a dipole magnet (DCT) with integrated quadrupole magnets (CCTs) housed within each side of the bore.
This integrated system provides both bending and focusing/defocusing capabilities for charged particles.
The DCT dipole (\SI{45}{\degree}) provides the primary bending, while the two CCT quadrupoles (\SI{200}{mrad}) achieve simultaneous focusing or defocusing.

For a pipe with a circular aperture of radius \(r=\SI{36}{mm}\) and central bend radius \(R=\SI{1}{m}\), the focal \(c\) in bipolar coordinates and the radial coordinate parameter \(\eta_0\) associated for the bore are given by:
\begin{equation}
	\begin{aligned}
		c &= \sqrt{R^2 - r^2}, \\
		\eta_0 &= \ln \left( \frac{R+r+\sqrt{R^2-r^2}}{R+r-\sqrt{R^2-r^2}} \right).
	\end{aligned}
	\label{eq:bipolar_coords}
\end{equation}

\begin{figure}[htbp]
	\centering
	\includegraphics[width=9.0cm]{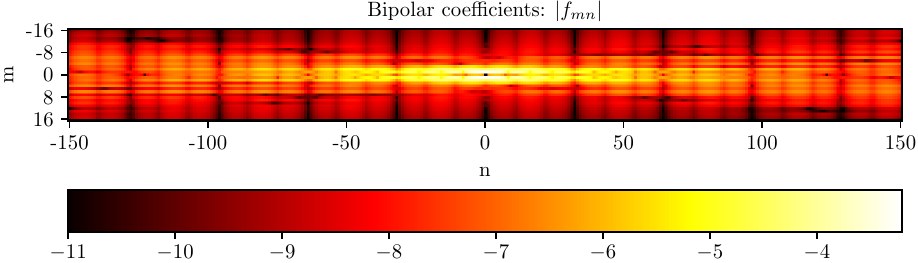}
	\caption{The bipolar coefficients \(f_{m,n}\), with higher retained orders of \(m\) and \(n\), lead to a more accurate representation of the magnetic field.}
	\label{fig:Bipolar_coefficients}
\end{figure}

The normal magnetic field \(B_\eta\) on the surface of bore \(\eta_0\) is used to obtain the bipolar coefficients \(f_{m,n}\).
Details are provided in \autoref{sec:cylindrical_appendix}~\cite{Vector_Potential_Ring}, and the results are presented in \autoref{fig:Bipolar_coefficients}.
The decay of color suggests that accurately representing the magnetic field of this system necessitates retaining \(n\) to a high order, although this significantly increases the computational demands of symplectic tracking.
In this work, we retain terms up to \(|n| < 60, |m| < 18\).

\begin{figure}[htbp]
	\centering
	\includegraphics[width=7cm]{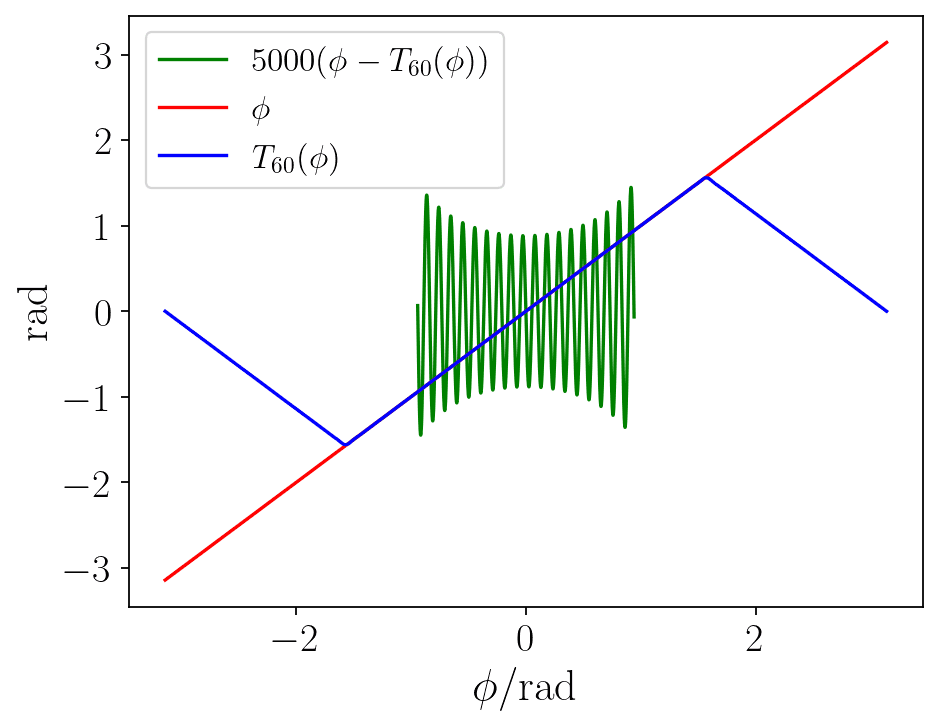}
	\caption{Approximation of the linear function \(\phi\) near the origin using a Fourier series. Given that the region of motion of interest is a segment near the origin, the Fourier series of a triangular wave (in blue) is employed instead of that of a sawtooth wave (in red). Higher truncation order reduces approximation errors (green).}
	\label{fig:ring_approximate_s}
\end{figure}

The vector potential in bipolar coordinates  by \autoref{vector_ring_potential} can be expressed as follows:
\begin{equation}
\begin{aligned}
	a_x(x, y, s) &= a_x^{(0)} s + \sum_{\substack{p=-\infty \\ p \neq 0}}^{\infty} a_x^{(p)} e^{\mathrm{i} k_p s}, \\
	a_y(x, y, s) &= a_y^{(0)} s + \sum_{\substack{p=-\infty \\ p \neq 0}}^{\infty} a_y^{(p)} e^{\mathrm{i} k_p s},
\end{aligned}
\end{equation}
where \( k_p = p/R \).
To express the longitudinal position (\(s\), or \(\phi\) ) in all functions in terms of trigonometric functions, we can employ a technique to handle the linear function associated with the longitudinal coordinate.
Without loss of generality, assume that \( \phi \) is symmetric around \( \phi = 0 \). As the segment we interest for particle tracking is around \( \phi = 0 \), we consider a Fourier decomposition of triangular wave and just use part of it, given by:
\begin{equation}
	\begin{aligned}
		&T_K\left( \phi \right) =\sum_{p=-K}^K{\frac{g(p)}{\text{i}p}e^{\text{i}p\phi}},\\
		&g(p)=0,\text{if\,\,}p=\text{even,}\\
		&g(p)=2\frac{\left( -1 \right) ^{\left( p-1 \right) /2}}{p\pi},\text{if\ }p=\text{odd}. 
		\label{trigonometric}
	\end{aligned}
\end{equation}
The use of a triangular wave instead of a direct Fourier transform for \(\phi\) is intended to facilitate faster convergence within the tracking interval, as shown \autoref{fig:ring_approximate_s}. 
As will be demonstrated later, this approximation simplifies the recurrence relationship.
The vector potential is changed as:
\begin{equation}
	\begin{aligned}
	a_x\left( x,y,s \right) &=\sum_{p=-K,p\ne 0}^K{\left( a_{x}^{\left( p \right)}+\frac{g\left( p \right)}{\text{i}p}a_{x}^{\left( 0 \right)} \right)}e^{\text{i}k_ps},\\
	a_y\left( x,y,s \right) &=\sum_{p=-K,p\ne 0}^K{\left( a_{y}^{\left( p \right)}+\frac{g\left( p \right)}{\text{i}p}a_{y}^{\left( 0 \right)} \right)}e^{\text{i}k_ps}.
	\end{aligned}
\end{equation}

We will switch back and forth between Cartesian and bipolar coordinates whenever it suits best.
The Hamilton-Jacobi equation for curved reference trajectories is given by
\begin{equation}
\begin{aligned}
	\partial _sF=\left( 1+hx \right) &\left[ 1-\frac{1}{2}\left( \partial _xF-\varepsilon a_x \right) ^2-\frac{1}{2}\left( \partial _yF-\varepsilon a_y \right) ^2 \right]\\
	=\left( 1+hx \right) &\left[ 1-\frac{1}{2}\left( \frac{k}{c}\left. \frac{\partial F}{\partial \eta} \right|_{\xi ,s}-\varepsilon a_{\eta} \right) ^2 \right.\\
	&\left. -\frac{1}{2}\left( \frac{k}{c}\left. \frac{\partial F}{\partial \xi} \right|_{\eta ,s}-\varepsilon a_{\xi} \right) ^2 \right] .\\
\end{aligned}
\end{equation}
Here the longitudinal vector potential \(a_s\) vanished due to the gauge we choose. 
And the functions \(a_\eta\) and \(a_\xi\) are the corresponding normalized potentials in bipolar coordinates, and the transformation relations are expressed as follows:
\begin{equation}
	\begin{aligned}
		a_{\eta} &= -a_x \left. \frac{\partial y}{\partial \xi} \right|_{\eta} + a_y \left. \frac{\partial x}{\partial \xi} \right|_{\eta}, \\
		a_{\xi} &= +a_x \left. \frac{\partial y}{\partial \eta} \right|_{\xi} - a_y \left. \frac{\partial x}{\partial \eta} \right|_{\xi}, \\
		a_x &= -a_{\eta} \left. \frac{\partial \xi}{\partial y} \right|_x + a_{\xi} \left. \frac{\partial \eta}{\partial y} \right|_x, \\
		a_y &= +a_{\eta} \left. \frac{\partial \xi}{\partial x} \right|_y - a_{\xi} \left. \frac{\partial \eta}{\partial x} \right|_y.
	\end{aligned}
\end{equation}
The momentum in bipolar coordinates is given by:
\begin{equation}
	\begin{aligned}
		p_{\eta} &=\frac{\partial F}{\partial \eta}=p_x\left. \frac{\partial x}{\partial \eta} \right|_{\xi}+p_y\left. \frac{\partial y}{\partial \eta} \right|_{\xi},\\
		p_{\xi} &=\frac{\partial F}{\partial \xi}=p_x\left. \frac{\partial x}{\partial \xi} \right|_{\eta}+p_y\left. \frac{\partial y}{\partial \xi} \right|_{\eta}.
	\end{aligned}
\end{equation}

Here, we introduce some Fourier decomposition of \(Z_{\left( ijkl \right)}\), \(X_{\left( ijkl \right)}\) and \(Y_{\left( ijkl \right)}\), as follows:
\begin{equation}
	\begin{aligned}
	&Z_{\left( ijkl \right)} := \frac{\partial f_{ijkl}}{\partial s} = \sum_{p=-\infty}^{\infty}{\mathbb{Z}_{\left( ijkl \right)}^{\left( p \right)}e^{\text{i}k_ps}},\\
	&X_{\left( ijkl \right)} := \frac{\partial f_{ijkl}}{\partial x}-\delta _{i}^{0}\delta _{j}^{0}\delta _{k}^{0}\delta _{l}^{1}a_x =\sum_{p=-\infty}^{\infty}{\mathbb{X}_{\left( ijkl \right)}^{\left( p \right)}e^{\text{i}k_ps}},\\
	&Y_{\left( ijkl \right)} := \frac{\partial f_{ijkl}}{\partial y}-\delta _{i}^{0}\delta _{j}^{0}\delta _{k}^{0}\delta _{l}^{1}a_y = \sum_{p=-\infty}^{\infty}{\mathbb{Y}_{\left( ijkl \right)}^{\left( p \right)}e^{\text{i}k_ps}}.
	\end{aligned}
\end{equation}
The coefficient of generating function \(f_{\left( ijkl \right)}\) can be reconstructed by \(Z_{\left( ijkl \right)}\) as follows:
\begin{equation}
	\begin{aligned}
		f_{\left( ijkl \right)}&=\mathbb{Z}_{\left( ijkl \right)}^{\left( 0 \right)}\left( s-s_f \right) +\\
		&\sum_{p=-\infty, p\neq 0}^{\infty}{\frac{e^{\text{i}k_ps}-e^{\text{i}k_ps_f}}{\text{i}k_p}\mathbb{Z}_{\left(ijkl \right)}^{\left( p \right)}},
	\end{aligned}
\end{equation}
where the condition \( \left. f_{\left( ijkl \right)} \right|_{s_f}\equiv 0 \) is satisfied automatically.
Bipolar coordinates can facilitate the deduction, as the variables are well separated and partial derivatives are simple. 
Here we introduce some partial derivatives of \(f_{ijkl}\) with respect to \(\eta\) and \(\xi\).
\begin{equation}
	\begin{aligned}
		f_{ijkl;\eta}-\frac{c}{k}\delta _{i}^{0}\delta _{j}^{0}\delta _{k}^{0}\delta _{l}^{1}a_{\eta}&=\sum_{p=-\infty ,p\ne 0}^{\infty}{e^{\text{i}k_ps}\mathcal{U}_{\left( ijkl \right)}^{\left( p \right)}},\\
		f_{ijkl;\xi}-\frac{c}{k}\delta _{i}^{0}\delta _{j}^{0}\delta _{k}^{0}\delta _{l}^{1}a_{\xi}&=\sum_{p=-\infty ,p\ne 0}^{\infty}{e^{\text{i}k_ps}\mathcal{V}_{\left( ijkl \right)}^{\left( p \right)}}.
	\end{aligned}
\end{equation}
For \((i, j, k, l ) \ne (0, 0, 0, 1)\), the approximation of \(\phi\) in  \autoref{trigonometric} results in:
\begin{equation}
	\begin{aligned}
		\mathcal{U}_{\left( ijkl \right)}^{\left( p \right)}&=\frac{\mathbb{Z}_{\left( ijkl \right)}^{\left( p;\eta \right)}+\mathbb{Z}_{\left( ijkl \right)}^{\left( 0;\eta \right)}g\left( p \right)}{\text{i}k_p},\\
		\mathcal{V}_{\left( ijkl \right)}^{\left( p \right)} &=\frac{\mathbb{Z}_{\left( ijkl \right)}^{\left( p;\xi \right)}+\mathbb{Z}_{\left( ijkl \right)}^{\left( 0;\xi \right)}g\left( p \right)}{\text{i}k_p},
	\end{aligned}p\ne 0.
\end{equation}
We then convert back to Cartesian coordinates as follows:
\begin{equation}
	\begin{aligned}
		\mathbb{X}_{\left( ijkl \right)}^{\left( p \right)} & =\left. \frac{\partial \eta}{\partial x} \right|_y\mathcal{U}_{\left( ijkl \right)}^{\left( p \right)}+\left. \frac{\partial \xi}{\partial x} \right|_y\mathcal{V}_{\left( ijkl \right)}^{\left( p \right)},\\
		\mathbb{Y}_{\left( ijkl \right)}^{\left( p \right)} & =\left. \frac{\partial \eta}{\partial y} \right|_x\mathcal{U}_{\left( ijkl \right)}^{\left( p \right)}+\left. \frac{\partial \xi}{\partial y} \right|_x\mathcal{V}_{\left( ijkl \right)}^{\left( p \right)},
	\end{aligned}p\ne 0.
\end{equation}
The term corresponding to \(p=0\) is derived as follows:
\begin{equation}
	\begin{aligned}
		\mathcal{U}_{\left( ijkl \right)}^{\left( 0 \right)}&=-\sum_{q=-\infty,  q\neq 0}^{\infty}{\mathcal{U}_{\left( ijkl \right)}^{\left( q \right)}}e^{\text{i}k_qs_f},\\
		\mathcal{V}_{\left( ijkl \right)}^{\left( 0 \right)} &=-\sum_{q=-\infty,  q\neq 0}^{\infty}{\mathcal{V}_{\left( ijkl \right)}^{\left( q \right)}}e^{\text{i}k_qs_f},
	\end{aligned}p=0.
\end{equation}
\begin{equation}
	\begin{aligned}
		\mathbb{X}_{\left( ijkl \right)}^{\left( 0 \right)} &=-\sum_{q=-\infty,  q\neq 0}^{\infty}{\mathbb{X}_{\left( ijkl \right)}^{\left( q \right)}}e^{\text{i}k_qs_f},\\
		\mathbb{Y}_{\left( ijkl \right)}^{\left( 0 \right)} &=-\sum_{q=-\infty, q\neq 0}^{\infty}{\mathbb{Y}_{\left( ijkl \right)}^{\left( q \right)}}e^{\text{i}k_qs_f},
	\end{aligned}p= 0.
\end{equation}
We obtain the recurrence relation for high-order coefficients of generating functions:
\begin{equation}
\begin{aligned}
	\mathbb{Z}_{\left( ijkl \right)}^{\left( p \right)}=&-\frac{1}{2}\frac{k^2}{c^2}\sum_{\alpha \beta \gamma \chi ,q} \Bigg[ \mathcal{U}_{\left( \alpha \beta \gamma \chi \right)}^{\left( q \right)}\mathcal{U}_{\left( i-\alpha ,j-\beta ,k-\gamma ,l-\chi \right)}^{\left( p-q \right)}  \\
	&+\mathcal{V}_{\left( \alpha \beta \gamma \chi \right)}^{\left( q \right)}\mathcal{V}_{\left( i-\alpha ,j-\beta ,k-\gamma ,l-\chi \right)}^{\left( p-q \right)} \Bigg]\\
	&-\frac{x}{2}\frac{k^2}{c^2}\sum_{\alpha \beta \gamma \chi ,q}\Bigg[ \mathcal{U}_{\left( \alpha \beta \gamma \chi \right)}^{\left( q \right)}\mathcal{U}_{\left( i-\alpha ,j-\beta ,k-1-\gamma ,l-\chi \right)}^{\left( p-q \right)}\\
	& +\mathcal{V}_{\left( \alpha \beta \gamma \chi \right)}^{\left( q \right)}\mathcal{V}_{\left( i-\alpha ,j-\beta ,k-1-\gamma ,l-\chi \right)}^{\left( p-q \right)} \Bigg].
\end{aligned}
\end{equation}
Finally, we calculate the coordinates and momentum at the end of the stepping.
\begin{equation}
	\begin{aligned}
		p_{xf} &= - \sum_{ijkl} (p_{xf})^i (p_{yf})^j h^k \varepsilon^l \left[  \mathbb{Z}_{(ijkl)}^{(0;x)} (s - s_f)    \right. \\
		& \left. + \sum_{p=-\infty, p \neq 0}^{\infty} \mathbb{Z}_{(ijkl)}^{(p;x)} \frac{e^{\text{i}ps_0} - e^{\text{i}ps_f}}{\text{i}p} \right] + p_{x0} , \\
		p_{yf} &= - \sum_{ijkl} (p_{xf})^i (p_{yf})^j h^k \varepsilon^l \left[  \mathbb{Z}_{(ijkl)}^{(0;x)} (s - s_f)    \right. \\
		& \left. + \sum_{p=-\infty , p \neq 0}^{\infty} \mathbb{Z}_{(ijkl)}^{(p;x)} \frac{e^{\text{i}ps_0} - e^{\text{i}ps_f}}{\text{i}p} \right] + p_{y0}.
	\end{aligned}
\end{equation}
\begin{equation}
	\begin{aligned}
		x_f =& \sum_{ijkl} i (p_{xf})^{i-1} (p_{yf})^j h^k \varepsilon^l \left[ \mathbb{Z}_{(ijkl)}^{(0)} (s_0 - s_f) \right. \\
		&\left. \sum_{p=-\infty,  p \neq 0}^{\infty} \mathbb{Z}_{(ijkl)}^{(p)} \frac{e^{\text{i}ps_0} - e^{\text{i}ps_f}}{\text{i}p}   \right]  + x_0, \\
		y_f =& \sum_{ijkl} j (p_{xf})^{i} (p_{yf})^{j-1} h^k \varepsilon^l \left[ \mathbb{Z}_{(ijkl)}^{(0)} (s_0 - s_f) \right. \\
		&\left. \sum_{p=-\infty,  p \neq 0}^{\infty} \mathbb{Z}_{(ijkl)}^{(p)} \frac{e^{\text{i}ps_0} - e^{\text{i}ps_f}}{\text{i}p}   \right] + y_0.
	\end{aligned}
\end{equation}

\begin{figure}[htbp]
	\centering
	\includegraphics[width=7cm]{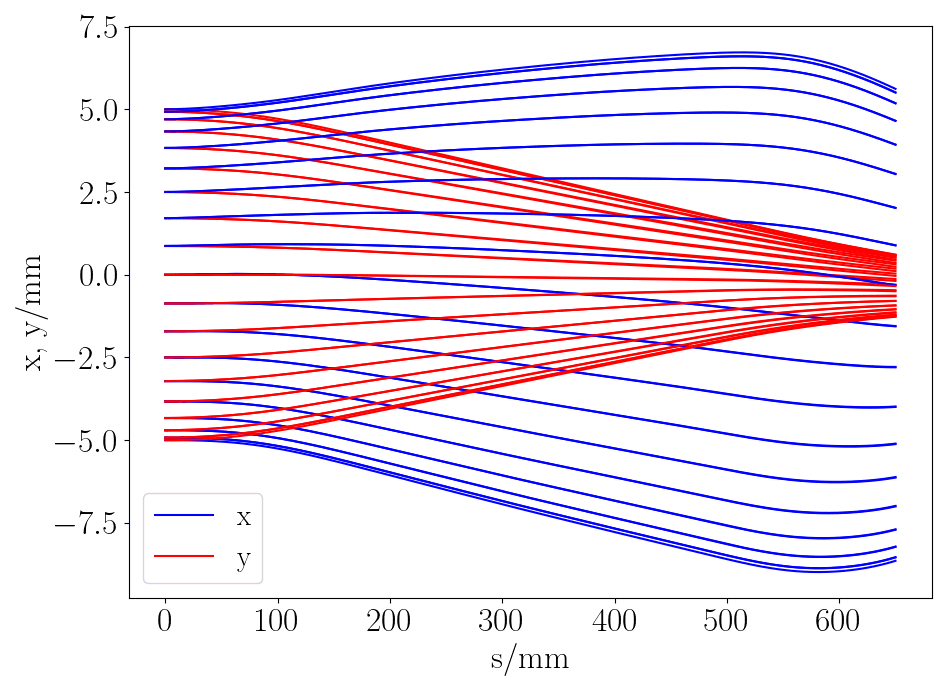}
	\caption{Symplectic tracking of particles in bipolar coordinates by generating function. The order of field harmonics is truncated to \(|n|\leq 60\) in generating function. }
	\label{fig:ring_envelope}
\end{figure}

\begin{figure}[htbp]
	\centering
	\includegraphics[width=7cm]{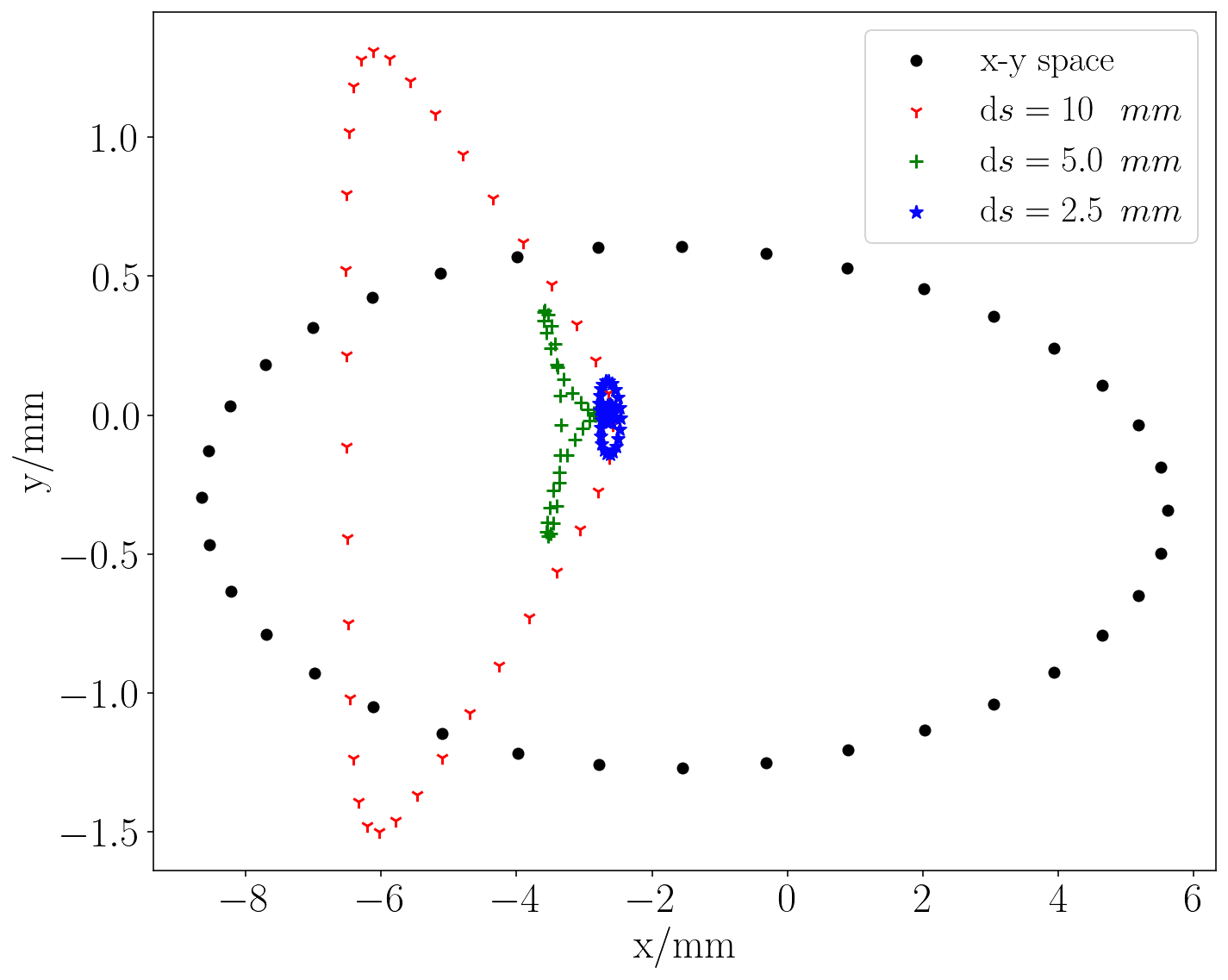}
	\caption{The tracking results in ring coordinates at the end of the beamline are shown  (\(i+j+k+l \le 2\)). The black point represents the results in \(x-y\) phase space. Three different step sizes are used for tracking through the beamline: red (\SI{10}{mm}), green (\SI{5}{mm}), and blue (\SI{2.5}{mm}). The difference between the GF and RK methods is amplified in the x-direction (\(\times 1000\)) and the y-direction (\(\times 500\)).}
	\label{fig:ring_tracking_results}
\end{figure}

The algorithm described above can be implemented through code.
We track 36 particles initially uniformly distributed within a circle of radius \SI{5}{mm}, observing their motion in the curved beamline shown in \autoref{fig:coil_displace}.
The results are presented in \autoref{fig:ring_envelope}.
The GF method is truncated at the order \(i+j+k+l \leq 2\).
Attempts to use higher-order truncations result in a substantial increase in memory requirements, making further calculations infeasible.
At the end of the beamline, we analyze the tracking results for different step sizes, \(\text{d} s= \SI{10}{mm}, \SI{5}{mm}, \SI{2.5}{mm}\) and compare them with the fine-step Runge-Kutta tracking results, as shown in \autoref{fig:ring_tracking_results}.
The error is amplified by a factor of 2000 in the x-direction and 500 in the y-direction for better comparison.
Overall, the error remains small and decreased as the step size is reduced.
Due to the Fourier expansion approximation of \(s\) and the finite expansion series of the generating function, the center of the error deviates from the origin even if the step size is small.
In general, for sufficiently ample computational resources and with high enough truncation of the expansion series in \(F\), symplectic tracking based on ring coordinates is feasible.
However, this approach is not practical on typical desktop computers.

\begin{figure}[htbp]
	\centering
	\includegraphics[width=6.0cm]{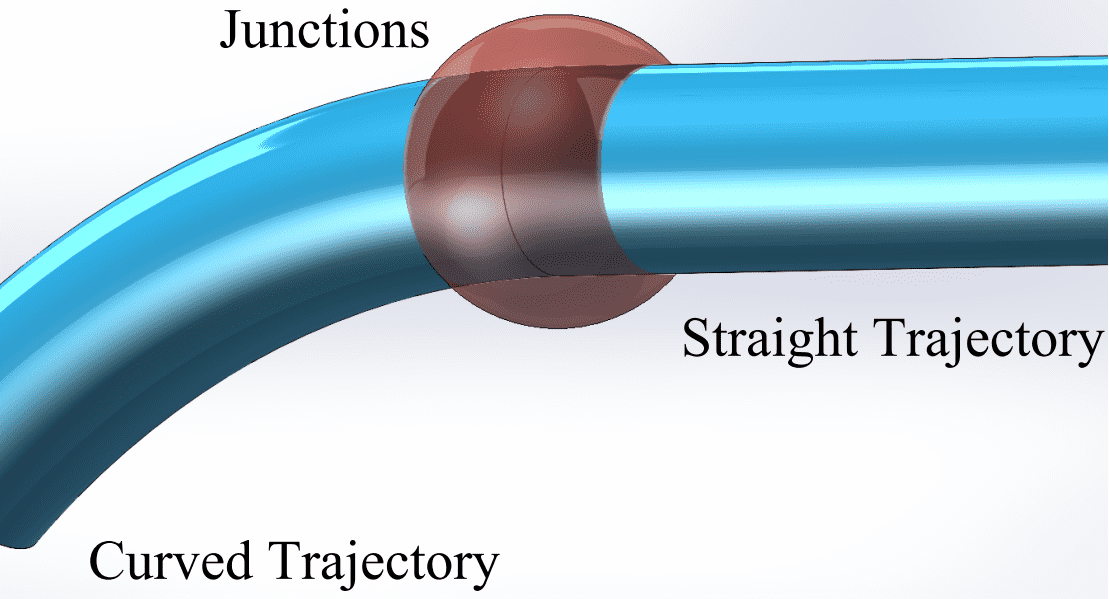}
	\caption{The beamline comprises both straight and curved segments. Cylindrical coordinates are applicable in the straight segments, while ring coordinates are more suitable for the curved segments. Spherical coordinates can be employed at the junctions due to the complex superposition of fields from two sides.}
	\label{fig:junctions_combined}
\end{figure}

Cylindrical coordinates are particularly well-suited for modeling straight beamlines, whereas ring coordinates are more appropriate for curved beamlines.
At the junctions where straight and curved reference beamlines meet, the superposition of magnetic fields from both sides necessitates a transition to spherical coordinates to accurately account for the field complexity, as depicted in \autoref{fig:junctions_combined}.
It is important to note that transitioning between different coordinate systems for particles requires a conversion between canonical and mechanical momentum.
Although this configuration can be used to track particles in any beamline, the approximation of longitudinal position through Fourier decomposition may introduce errors over long-term tracking.
Additionally, the high-order expansion of the generating function demands substantial computational resources, as the field representation in both cylindrical and ring coordinates involves numerous terms.
In the following subsection, we will introduce an alternative method.

\subsection{\label{sec:Tracking_sphere}Tracking through a series of spheres}

\begin{figure}[htbp]
	\centering
	\includegraphics[width=6.5cm]{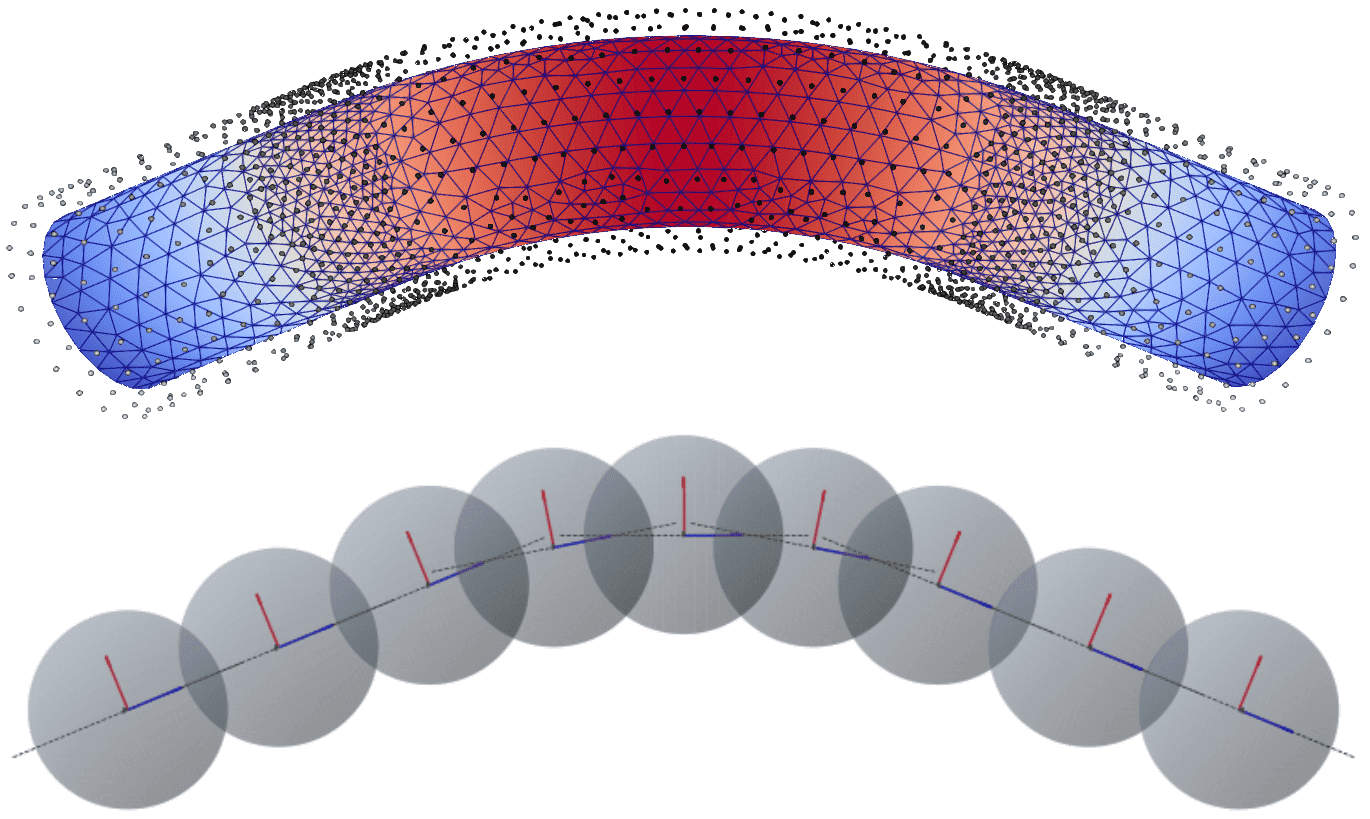}
	\caption{By placing magnetic moments or monopoles externally (represented as the surrounding dots), outside the pipe, to satisfy the specified normal magnetic field boundary conditions \(B_n\) on its surface, the magnetic field within the surface is uniquely determined. The reconstructed field is stored in a series of spheres, each with its own Cartesian coordinate system. As particles move, their coordinate system and sphere index should be adjusted accordingly. Arbitrary beamlines can be simulated through the assembly of multiple spheres.}
	\label{fig:sphere_junctions}
\end{figure}

The uniqueness theorem in electromagnetism states that if the normal component of the magnetic field \(B_n\) is known on a closed surface within a source-free region, then the magnetic field inside is uniquely determined.
Furthermore, the two fundamental sources that generate magnetic fields are magnetic moments, which correspond to current loops and can also be interpreted as magnetic dipoles, and hypothetical magnetic monopoles.
We can artificially place magnetic moments or monopoles externally, outside the region of interest, to achieve the same normal magnetic field boundary conditions \(B_n\) (called Neumann boundary condition) on the surface of the region used for tracking, thereby enabling the reconstruction of the magnetic field within the surface.
The magnetic field generated by both magnetic dipoles and monopoles can be expressed in terms of spherical harmonics, as detailed in the \autoref{sec:sphere_appendix}.
A key advantage of this representation is its expressibility in Cartesian coordinates transformed from spherical coordinates.
For a beamline composed of straight and curved segments connected in series, a sequence of spherical elements can be used to model such a beamline.
Each sphere in this sequence corresponds to a local Cartesian coordinate system.
Within each sphere, local magnetic field information is encoded using spherical harmonics calculated from the reconstructed sources, as seen in \autoref{fig:sphere_junctions}.

\begin{figure}[htbp]
	\centering
	\includegraphics[width=6.0cm]{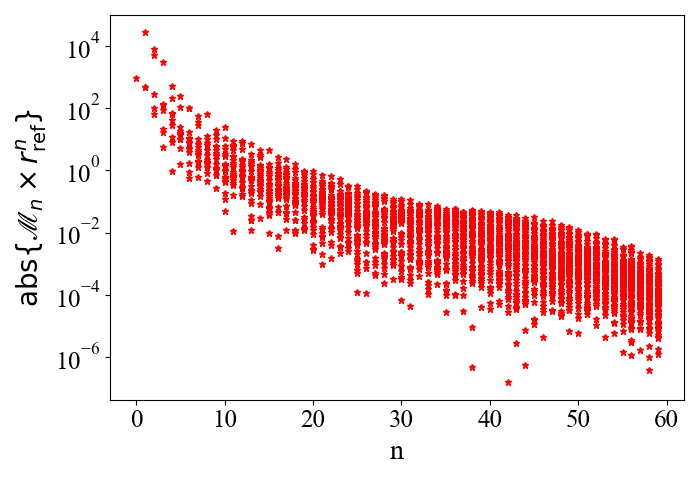}
	\caption{The strength of field harmonics in one sphere along the beamline. The strength of the field harmonics decreases rapidly with increasing order of spherical harmonics, and very high-order harmonics can be neglected for field map.}
	\label{fig:scalar_potential_fig}
\end{figure}

The beamline shown in \autoref{fig:coil_displace} consists of curved segments and extended straight sections.
Magnetic dipoles are employed to reconstruct the magnetic fields, and the details of procedure are shown in~\cite{JieLi_Reconstruction}.
And the strength of the field harmonics used to represent the scalar potential in a sphere along the beamline is illustrated in \autoref{fig:scalar_potential_fig}.
The vector potential can be computed using the relation in \autoref{scalar_vector_straight}.
High-order harmonics can be neglected in the field representation due to the rapid decay of the strength.
The advantage of this approach is the high precision of magnetic field reconstruction, allowing for an analytical representation of the local magnetic field, including the magnetic vector potential and magnetic scalar potential.
The disadvantage is that, as the beam moves, it is necessary to switch between different spherical coordinate systems based on its position.
The particle's momentum switches between canonical and mechanical momentum.

The Hamilton-Jacobi equation becomes simpler within spherical regions that are described by Cartesian fields, as opposed to curved coordinates, due to a reduction in the degrees of freedom. The simplified form of the equation is given by:
\begin{equation}
	\partial _zF= 1-\frac{1}{2}\left( 	\partial _x F - \varepsilon a_{x} \right) ^2-\frac{1}{2}\left( 	\partial _y F-\varepsilon a_{y} \right) ^2 .
\end{equation}
The longitudinal coordinate is represented within the coefficients of the generating function  by means of Taylor series rather than trigonometric functions.
The coefficients \(f_{\left( ijl \right)}\) in the generating function can be decomposed as follows:
\begin{equation}
	\begin{aligned}
		Z_{\left( ijl \right)}&:=\frac{\partial f_{\left( ijl \right)}}{\partial z}=\sum_{p=0}^{\infty}{\mathbb{Z}_{\left( ijl \right)}^{\left( p \right)}z^p},\\
		f_{\left( ijl \right)}&=\sum_{p=1}^{\infty}{\mathbb{Z}_{\left( ijl \right)}^{\left( p-1 \right)}\frac{z^p-z_{f}^{p}}{p}}.\\
	\end{aligned}
\end{equation}
The derivatives of \(f_{\left( ijl \right)}\) with respect to \(x, y\) are represented as follows:
\begin{equation}
	\begin{aligned}
		X_{\left( ijl \right)}:=\frac{\partial f_{\left( ijl \right)}}{\partial x} - \delta _{i}^{0}\delta _{j}^{0}\delta _{l}^{1} a_x =&\sum_{p=0}^{\infty}{\mathbb{X}_{\left( ijl \right)}^{\left( p \right)}z^p},\\
		Y_{\left( ijl \right)}:=\frac{\partial f_{\left( ijl \right)}}{\partial y} - \delta _{i}^{0}\delta _{j}^{0}\delta _{l}^{1} a_y=&\sum_{p=0}^{\infty}{\mathbb{Y}_{\left( ijl \right)}^{\left( p \right)}z^p.}
	\end{aligned}
\end{equation}
For \((i,j,l) \ne (0, 0, 1)\), we have:
\begin{equation}
	\mathbb{X}_{(ijl)}^{(p)} = 
	\left\{
	\begin{array}{@{}ll}
		\frac{\mathbb{Z}_{(ijl)}^{(p-1;x)}}{p}, & \text{if } p > 0; \\[10pt]
		-\sum_{q=1}^{\infty} \mathbb{X}_{(ijl)}^{(q)} z_f^q, & \text{if } p = 0.
	\end{array}
	\right.
\end{equation}
\begin{equation}
	\mathbb{Y}_{(ijl)}^{(p)} = 
	\begin{cases} 
		\frac{\mathbb{Z}_{(ijl)}^{(p-1;y)}}{p}, & \text{if } p > 0; \\[10pt]
		-\sum_{q=1}^{\infty} \mathbb{Y}_{(ijl)}^{(q)} z_f^q, & \text{if } p = 0.
	\end{cases}
\end{equation}
The recurrence relation for high-order coefficients is given by:
\begin{equation}
	\begin{aligned}
		\mathbb{Z}_{(ijl)}^{(p)} &= -\frac{1}{2} \sum_{(\alpha, \beta, \chi) } \sum_{q=0}^{p} \Bigg[ \mathbb{X}_{(\alpha, \beta, \chi)}^{(q)} \mathbb{X}_{(i-\alpha, j-\beta, l-\chi)}^{(p-q)} \\
		&\quad + \mathbb{Y}_{(\alpha, \beta, \chi)}^{(q)} \mathbb{Y}_{(i-\alpha, j-\beta, l-\chi)}^{(p-q)} \Bigg].
	\end{aligned}
\end{equation}
The Newton routine to obtain the final momentum is given by.
\begin{equation}
	\begin{aligned}
		p_{xf} =& p_{x0} -\\
		& \sum_{ijl} \left[ \left( \sum_{p=1}^{\infty} \mathbb{Z}_{(ijl)}^{(p-1;x)} \frac{z_0^p - z_f^p}{p} \right) (p_{xf})^i (p_{yf})^j \varepsilon^l \right], \\
		p_{yf} =& p_{y0} -\\
		& \sum_{ijl} \left[ \left( \sum_{p=1}^{\infty} \mathbb{Z}_{(ijl)}^{(p-1;y)} \frac{z_0^p - z_f^p}{p} \right) (p_{xf})^i (p_{yf})^j \varepsilon^l \right].
	\end{aligned}
\end{equation}
Then the final coordinate is derived:
\begin{equation}
	\begin{aligned}
		x_f &= \sum_{ijl} \left( \sum_{p=1}^{\infty} \mathbb{Z}_{(ijl)}^{(p-1)} \frac{z_0^p - z_f^p}{p} \right) i (p_{xf})^{i-1} (p_{yf})^j \varepsilon^l + x_0, \\
		y_f &= \sum_{ijl} \left( \sum_{p=1}^{\infty} \mathbb{Z}_{(ijl)}^{(p-1)} \frac{z_0^p - z_f^p}{p} \right) (p_{xf})^i j (p_{yf})^{j-1} \varepsilon^l + y_0.
	\end{aligned}
\end{equation}

\begin{figure}[htbp]
	\centering
	\includegraphics[width=7cm]{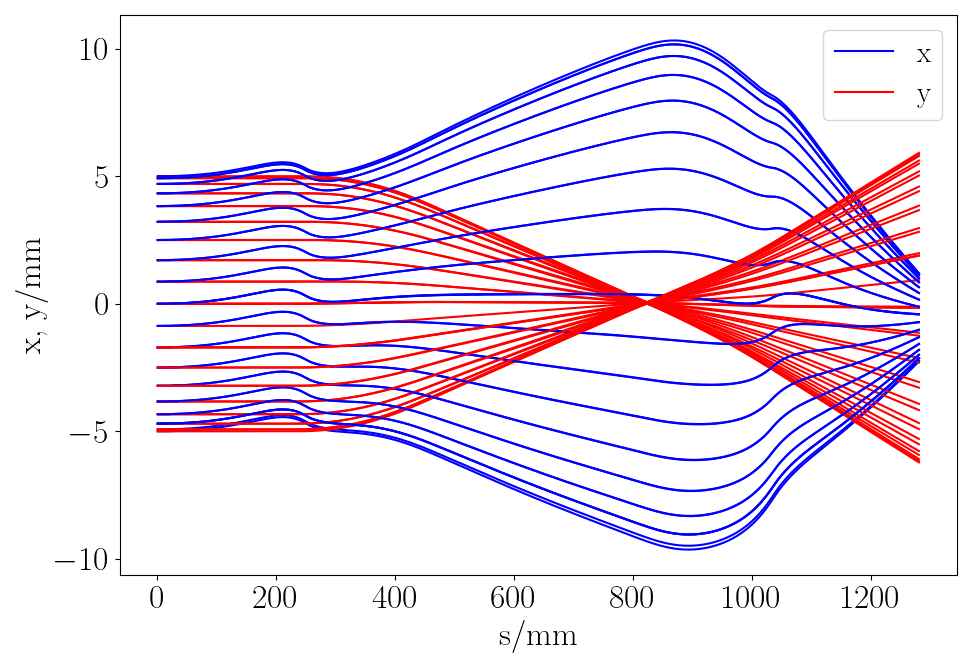}
	\caption{Symplectic tracking of particles in spherical coordinates by generating function.}
	\label{fig:sphere_envelope}
\end{figure}

\begin{figure}[htbp]
	\centering
	\includegraphics[width=7cm]{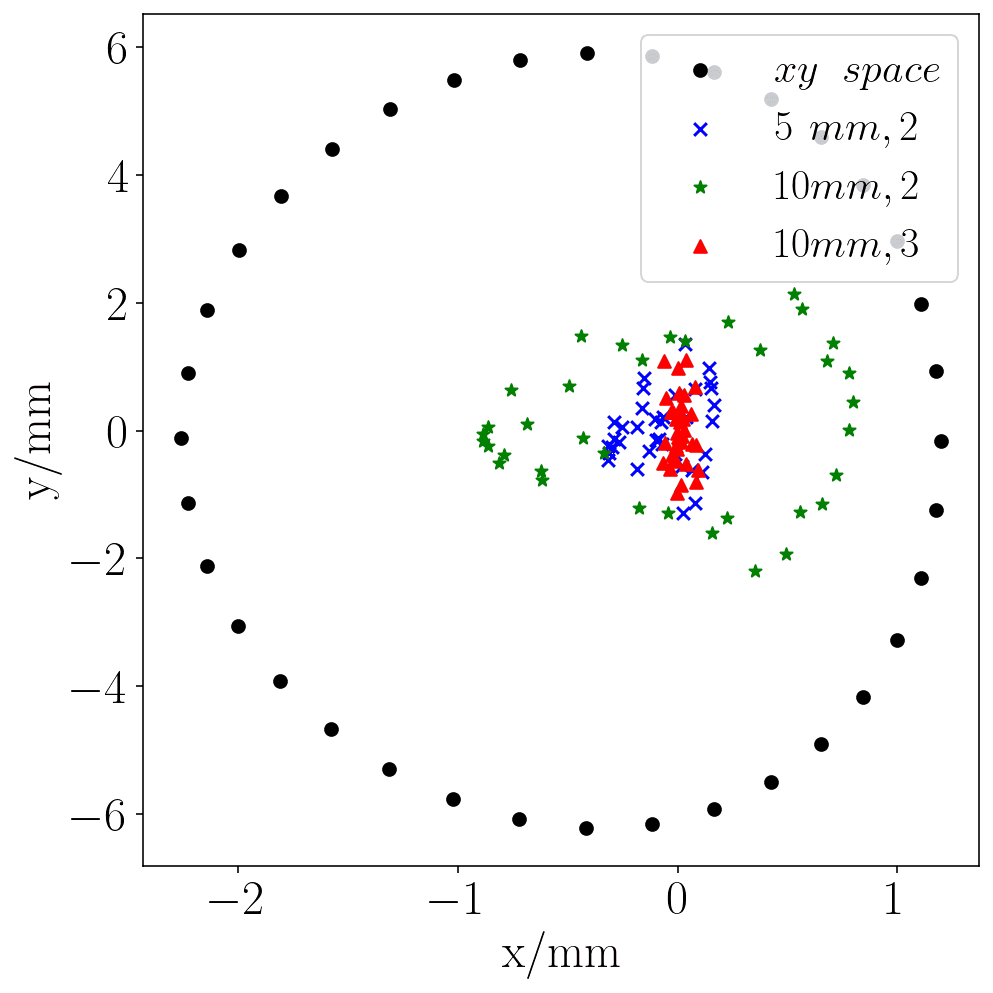}
	\caption{The tracking results in spherical coordinates at the end of the beamline are shown. The black point represents the results in the \(x-y\) phase space. Three different settings for the GF method are used for tracking through the beamline: blue (\(\text{d}s = \SI{5}{mm}\), order truncated to \(i+j+l \le 2\)), green (\(\text{d}s = \SI{10}{mm}\), order truncated to \(i+j+l \le 2\)), and red (\(\text{d}s = \SI{10}{mm}\), order truncated to \(i+j+l \le 3\)). The difference between the GF and RK methods is amplified in the x-direction (\(\times 1000\)) and the y-direction (\(\times 1000\)).}
	\label{fig:sphere_tracking_results}
\end{figure}

The spherical harmonics used to represent the field are truncated at order \(p \le 50\) to ensure precise reconstruction of the field.
We track 36 particles initially uniformly distributed within a circle of radius \SI{5}{mm} in the beamline (including both straight and curved segments), and the results are presented in \autoref{fig:sphere_envelope}.
At the end of the beamline, we analyze the tracking results for different settings, including varied step sizes \(\text{d} s\) and truncated orders for the generating function \(i+j+l\), and compare them with the fine-step Runge-Kutta tracking results, as shown in \autoref{fig:sphere_tracking_results}.
The error is amplified by a factor of 1000 in the x-direction and 1000 in the y-direction for better comparison.
Overall, the error is very small.
Even with a step size as large as \SI{10}{mm}, truncating the order to \(i+j+l \le 3\) still yields excellent tracking results.
Moreover, the computational effort is acceptable.

\section{\label{sec:conclu}Conclusion}

We introduce a tracking scheme based on the generating function for curved magnetic elements.
The field description is detailed under axisymmetric conditions, or through 3D harmonic expansions.
The Hamilton-Jacobi equation is solved using the perturbation method. Several tests have been conducted to validate the algorithm, including hard-edge magnets under axisymmetric conditions, curved magnets with a reference orbit described by toroidal(ring) functions, and multi-sphere models for any beamline.

It is noted that the accuracy of this calculation is dependent on the orders of expansions for generating function \(F\), and, in principle, can be enhanced to achieve arbitrary precision given sufficient computational resources.
The new method is expected to be extremely useful for beamline with special magnet designs, such as those employed in proton therapy with compact superconducting magnets, or in intricate field configurations within ring accelerators.

\begin{acknowledgments}
This work was supported by the National Natural Science Foundation of China (Grants No.12105005, No.12205007), and the National Grand Instrument Project (No.2019YFF01014403).
\end{acknowledgments}


\appendix

\section{\label{sec:cylindrical_appendix}Field representation in cylindrical coordinates}

The two referenced papers, \cite{Accurate_transfer} and \cite{Vector_Potential_Ring}, describe the procedure for extracting the relevant field representation coefficients for harmonic functions using cylindrical or ring surface data.
For both straight and curved beamline elements, a multipole decomposition approach, based on fitting numerical data to a bounding surface, ensures a significant reduction of residuals within the bounded region.

Within an infinitely extended straight cylindrical geometry, the magnetic scalar potential can be expressed as:
\begin{equation}
	\begin{aligned}
	\psi \left( \rho ,\theta ,z \right) &=\sum_{m\ge 0}{\int_{-\infty}^{\infty}{\text{d}k\times \psi ^{\left( m \right)}\left( k \right) I_m\left( k\rho \right) e^{\text{i}m\theta}e^{\text{i}kz}}},
	\end{aligned}
\end{equation}
where \(I_m\left( k\rho \right)\) means the modified Bessel function~\cite{tuprints11687}.
For a cylindrical geometry with longitudinal periodicity \(L\), the magnetic scalar potential is shown as follows:
\begin{equation}
	\begin{aligned}
		\psi \left( \rho ,\theta ,z \right) =&\sum_{m\ge 0}{\psi ^{\left( m,0 \right)}\rho ^me^{\text{i}m\theta}} + \\
		&\sum_{m\ge 0}{\sum_{k\ne 0}{\psi ^{\left( m,k \right)}I_m\left( k\frac{2\pi}{L}\rho \right) e^{\text{i}m\theta}e^{\text{i}k\frac{2\pi}{L}z}}}.\\
	\end{aligned}
\end{equation}
The radial magnetic field  on the cylindrical surface is calculated as follows: 
\begin{equation}
	\begin{aligned}
	B_{\rho} =&\sum_{m\ge 0}{\psi ^{\left( m,0 \right)}m\rho ^{m-1}e^{\text{i}m\theta}}+\\
	&\sum_{m\ge 0}{\sum_{k\ne 0}{\psi ^{\left( m,k \right)}\frac{\partial I_m}{\partial \rho}e^{\text{i}m\theta}e^{\text{i}k\frac{2\pi}{L}z}}}.\\
	\end{aligned}
\end{equation}
In the magnet simulation software, the coefficients \(\psi^{(m,k)}\) can be derived from \(B_{\rho}\) by performing two-dimensional Fourier transforms separately with respect to \(\theta\) and \(z\).

Similarly, the ``radial'' magnetic field \(B_{\eta}\) on the bore surface of a ring geometry is shown as follows:
\begin{equation}
	\begin{aligned}
		B_{\eta}(\eta_0, \xi, \phi) &= \frac{\text{k}^{\frac{3}{2}}}{c} \sum_{m=-\infty}^{\infty} \sum_{n=-\infty}^{\infty} f_{m,n} e^{\text{i}n\phi} e^{\text{i}m\xi} \times \\
		&\left\{ \frac{\sinh(\eta_0)}{2\text{k}}  + \frac{\frac{\partial}{\partial \eta}Q_{|m|-\frac{1}{2}}^{|n|}(\cosh(\eta_0))}{Q_{|m|-\frac{1}{2}}^{|n|}(\cosh(\eta_0))} \right\}.
	\end{aligned}
\end{equation}
Here, the ``angle'' variable \(\xi\) is not well separated in the expression.
To determine the coefficients \(f_{m,n}\), we employ a simple analytical approach.
The Fourier decomposition on the ring surface \(\eta_0\) with respect to the geometric angle \(\theta\) is straightforward to derive. Concurrently, the Fourier decomposition of each ring harmonic in terms of \(\theta\) can also be numerically obtained.
By establishing a comparative framework between these two decompositions, we formulate a matrix equation that uniquely determines the coefficients \(f_{m,n}\).

\section{\label{sec:sphere_appendix}Field generated by the magnetic moments}
The Helmholtz theorem states that the inner fields can be reconstructed by the magnetic field on the surface of the region~\cite{Manikonda2006ANAH}:
\begin{equation}
	\begin{aligned}
		\vec{B}\left( \vec{x} \right) &=\nabla\times \vec{A}_t\left( \vec{x} \right) +\nabla\varphi _n\left( \vec{x} \right),\\
		\varphi _n\left( \vec{x} \right) &=\frac{1}{4\pi}\int_{\partial \Omega}{\frac{\vec{n}\left( \vec{x}_{\text{s}} \right) \cdot \overrightarrow{B}\left( \vec{x}_{\text{s}} \right)}{|\vec{x}-\vec{x}_{\text{s}}|}}\text{d}s,\\
		\vec{A}_t\left( \vec{x} \right) &=-\frac{1}{4\pi}\int_{\partial \Omega}{\frac{\vec{n}\left( \vec{x}_{\text{s}} \right) \times \vec{B}\left( \vec{x}_{\text{s}} \right)}{|\vec{x}-\vec{x}_{\text{s}}|}}\text{d}s.	
	\end{aligned}
\end{equation}
Moreover, the uniqueness theorem in electrodynamics posits that the normal component of the magnetic field \(B_n\) on the boundary surface 
of the region is sufficient to uniquely determine and reconstruct the internal magnetic field.
If the magnetic moments or monopoles outside the region are appropriately configured to satisfy the \(B_n\) boundary condition, the internal magnetic field can be accurately reconstructed.

Considering a cluster of current elements \(I_j\text{d}\vec{l}_j\) with spherical coordinates \((r_j, \theta_j, \phi_j)\) far away from the origin, the vector potential at the origin can be calculated by the Biot-Savart law as follows:
\begin{equation}
	\begin{aligned}
		\vec{A}\left( r,\theta ,\phi \right) &=\mu _0\sum_{n=0}^{\infty}{\sum_{m=-n}^n{\vec{M}_{n,m}\times r^nY_{n,m}\left( \theta ,\phi \right)}},\\
		\vec{M}_{n,m}&=\sum_j{\frac{I_j\text{d}\vec{l}_j}{2n+1}}\frac{Y_{n,m}\left( \theta _j,\phi _j \right)}{r_{j}^{n+1}}.
	\end{aligned}
	\label{eq:sphere_formula}
\end{equation}
The magnetic moment can be defined as the limit of a current loop.
The coefficient \( \vec{M}_{n,m} \) for a magnetic moment located at \((r_0, \theta_0, \phi_0)\) with direction \((\Theta, \Phi)\) is calculated as follows: 
\begin{equation}
	\begin{aligned}
		\vec{M}_{n,m} = & \frac{|\vec{m}|(n+1)}{r_{0}^{n+2}(2n+1)}Y_{n,m} \left( \hat{r}_0 \times \hat{m} \right)+ \\
		&  \frac{|\vec{m}|}{r_{0}^{n+2}(2n+1)} \frac{\partial Y_{n,m}}{\partial \cos \theta} \left(-\hat{z} \times \hat{m} - \frac{z_0}{r_0} \hat{m} \times \hat{r}_0 \right) + \\
		& \frac{|\vec{m}|}{r_{0}^{n+2}\left( 2n+1 \right)}\frac{\partial Y_{n,m}}{\partial \phi}\left[ \frac{\cos \Theta \sin \left( \Phi -\phi _0 \right)}{\sin \theta _0}\frac{\hat{z}\times \hat{m}}{|\hat{z}\times \hat{m}|} - \right. \\
		& \left. \frac{\cos \left( \Phi -\phi _0 \right)}{\sin \theta _0}\frac{\left( \hat{z}\cdot \hat{m} \right) \hat{m}-\hat{z}}{|\hat{z}\times \vec{m}|} \right] .
	\end{aligned}
\end{equation}
Specially, if the magnetic moment is located at the polar axis \(\theta_0 = 0, \pi\), the only non-zero coefficients \( \vec{M}_{n,m} \) are shown:
\begin{equation}
	\begin{aligned}
		&\vec{M}_{n}^{-1}=-\cos^{n+1}\theta_0 \frac{-m_x\hat{e}_z+m_z\hat{e}_x}{2n+1}\frac{K_{n}^{1}}{r_{0}^{n+2}}\frac{n\left( n+1 \right)}{\sqrt{2}},\\
		&\vec{M}_{n}^{0}=\cos^{n+1}\theta_0 \frac{m_x\hat{e}_y-m_y\hat{e}_x}{2n+1}\frac{\left( n+1 \right) K_{n}^{0}}{r_{0}^{n+2}},\\
		&\vec{M}_{n}^{1}=-\cos^{n+1}\theta_0 \frac{m_y\hat{e}_z-m_z\hat{e}_y}{2n+1}\frac{K_{n}^{1}}{r_{0}^{n+2}}\frac{n\left( n+1 \right)}{\sqrt{2}}.
	\end{aligned}
\end{equation}

The scalar potential of a magnetic moment can be represented by spherical harmonics as follows:
\begin{equation}
	\begin{aligned}
		\varphi &=\mu _0\sum_{n=0}^{\infty}{\sum_{m=-n}^n{r^nY_{n,m}\left( \theta ,\phi \right) \mathscr{M}_{n,m}}},\\
		\mathscr{M}_{n,m}&=\sum_{n=0}^{\infty}{\sum_{m=-n}^n{\frac{|\vec{m}|}{\left( 2n+1 \right) r_{0}^{n+2}}\left[ \left( n+1 \right)  \hat{r}_0\cdot \hat{m}  Y_{n,m}  \right.}} + \\
		&\left. \frac{\partial Y_{n,m}}{\partial \cos \theta}\left( \frac{z_0}{r_0}\hat{r}_0\cdot \hat{m}-\cos \Theta \right) -\frac{\partial Y_{n,m}}{\partial \phi}\frac{\hat{z}\cdot \left( \hat{r}_0\times \hat{m} \right)}{\left( \hat{r}_0\times \hat{z} \right) ^2} \right].\\
	\end{aligned}
\end{equation}
Specially, if the magnetic moment is located at the polar axis \(\theta_0 = 0, \pi\), the only non-zero coefficients \( \mathscr{M}_{n,m} \) are shown:
\begin{equation}
	\begin{aligned}
		\mathscr{M}_{n,-1} &= -
		\cos^{n+1} \theta _0 \frac{K_{n}^{1}m_y}{\sqrt{2}(2n+1)}\frac{\left( n+1 \right) n}{r_{0}^{n+2}},\\
		\mathscr{M}_{n, 0} &= \cos^{n+1} \theta _0  \frac{K_{n}^{0}m_z}{2n+1}\frac{n+1}{r_{0}^{n+2}},\\
		\mathscr{M}_{n, 1} &= - \cos^{n+1} \theta _0  \frac{K_{n}^{1}m_x}{\sqrt{2}(2n+1)}\frac{\left( n+1 \right) n}{r_{0}^{n+2}}.
	\end{aligned}
\end{equation}

The expression of spherical harmonics can be presented by Legendre functions as follows:
\begin{equation}
	Y_{n,m}\left( \theta ,\phi \right) =\begin{cases}
		\sqrt{2}K_{n}^{m}\cos \left( m\phi \right) P_{n}^{m}\left( \cos \theta \right) (-1)^m ,&		m>0,\\
		\sqrt{2}K_{n}^{m}\sin \left( -m\phi \right) P_{n}^{-m}\left( \cos \theta \right)(-1)^m ,&		m<0,\\
		K_{n}^{0}P_{n}^{0}\left( \cos \theta \right)  ,&		m=0.\\
	\end{cases}
	\label{eq:shpere_how_cal}
\end{equation}
\begin{equation}
	K_{n}^{m}=\sqrt{\frac{\left( 2n+1 \right)}{4\pi}\frac{\left( n-\left| m \right| \right) !}{\left( n+\left| m \right| \right) !}}.
	\label{eq:shpere_how_cal_K}
\end{equation}
Switching from spherical to Cartesian coordinates, the formula for spherical harmonics is given by:
\begin{equation}
	\begin{aligned}
		r^nY_{n,m}=&\sqrt{\frac{2n+1}{2\pi}}\sum _{k=0}^{\left( n-m \right) //2}\sum _{r=0}^{k}\sum _{s=0}^{k-r}\sum _{p=0}^{m} \times \\
		&f_{k,r,s,p}^{n,m} x^{p+2r}y^{-p+2s+m}z^{n-2r-2s-m},\\
		f_{k,r,s,p}^{n,m}=& \left[
		\left( -1 \right) ^k\left( 2n-2k \right) !\cos \left( \frac{\pi}{2}\left( m-p \right) \right) C_{m}^{p} \right] \\
		&/ \left[ 2^n r!s!\left( k-r-s \right) !\left( n-k \right) !\left( n-2k-m \right) !m! \right]\\
		&/\left[\sqrt{\left( n+m \right) !/\left( n-m \right) !}/m!\right],
	\end{aligned}
	\label{eq:new_express_sphere}
\end{equation}
which is suitable for straight magnetic elements.

\newpage

\nocite{*}

%

\end{document}